\documentclass[english]{IEEEtran}
\usepackage[T1]{fontenc}
\usepackage[latin9]{inputenc}
\usepackage{color}
\usepackage{babel}
\usepackage{amsmath}
\usepackage{amsthm}
\usepackage{amssymb}
\usepackage{graphicx}
\makeatletter

\usepackage{mathtools}
\mathtoolsset{showonlyrefs}

\providecommand{\tabularnewline}{\\}

\theoremstyle{plain}
\newtheorem{thm}{\protect\theoremname}
\theoremstyle{plain}
\newtheorem{conjecture}{\protect\conjecturename}
\theoremstyle{plain}
\newtheorem{cor}{\protect\corollaryname}
\theoremstyle{remark}
\newtheorem{rem}{\protect\remarkname}
\theoremstyle{plain}
\newtheorem{lem}{\protect\lemmaname}
\theoremstyle{definition}
\newtheorem{defn}{\protect\definitionname}

\usepackage{babel}
\usepackage{accents} 
\usepackage{enumerate}
\usepackage{verbatim}
\usepackage{mathrsfs}
\usepackage{cite}
\usepackage{tikz}
\usepackage{bm,bbm}
\usepackage{lineno}
\newcommand{\bone}{\mathbbm{1}}

\usepackage[T3,T1]{fontenc}
\DeclareSymbolFont{tipa}{T3}{cmr}{m}{n}
\DeclareMathAccent{\invbreve}{\mathalpha}{tipa}{16}

\def\undertilde#1{\mathord{\vtop{\ialign{##\crcr
$\hfil\displaystyle{#1}\hfil$\crcr\noalign{\kern1.5pt\nointerlineskip}
$\hfil\tilde{}\hfil$\crcr\noalign{\kern1.5pt}}}}}

\newcommand{\subsetsim}{\mathrel{%
  \ooalign{\raise0.2ex\hbox{$\subset$}\cr\hidewidth\raise-0.8ex\hbox{\scalebox{0.9}{$\sim$}}\hidewidth\cr}}}
\newcommand{\supsetsim}{\mathrel{%
  \ooalign{\raise0.2ex\hbox{$\supset$}\cr\hidewidth\raise-0.8ex\hbox{\scalebox{0.9}{$\sim$}}\hidewidth\cr}}}

\newcommand{\subsetapprox}{\mathrel{%
  \ooalign{\raise0.4ex\hbox{$\subset$}\cr\hidewidth\raise-0.8ex\hbox{\scalebox{0.9}{$\approx$}}\hidewidth\cr}}}
\allowdisplaybreaks[1]
\newcommand{\calC}{\mathcal{C}}

\newcommand{\calP}{\mathcal{P}}

\newcommand{\calX}{\mathcal{X}}
\newcommand{\calY}{\mathcal{Y}}

\def\BSC{\mathrm{BSC}}
\def\DSBS{\mathrm{DSBS}}

\def\conv{\mathrm{conv}}
\def\conc{\mathrm{conc}}
\def\D{\mathsf{D}}
\def\tr{{\rm trace}}

\def\1{\mathbf{1}}

\providecommand{\corollaryname}{Corollary}
\providecommand{\lemmaname}{Lemma}

\providecommand{\remarkname}{Remark}
\providecommand{\theoremname}{Theorem}

\makeatother

\providecommand{\conjecturename}{Conjecture}
\providecommand{\corollaryname}{Corollary}
\providecommand{\definitionname}{Definition}
\providecommand{\lemmaname}{Lemma}
\providecommand{\remarkname}{Remark}
\providecommand{\theoremname}{Theorem}

\begin{document}
\title{Gray--Wyner and Mutual Information Regions for Doubly Symmetric
Binary Sources and Gaussian Sources }
\author{Lei Yu 
\thanks{L. Yu is with the School of Statistics and Data Science, LPMC, KLMDASR,
and LEBPS, Nankai University, Tianjin 300071, China (e-mail: leiyu@nankai.edu.cn).
This work was supported by the NSFC under grant 62101286 and the Fundamental
Research Funds for the Central Universities of China (Nankai University)
under grant 054-63233073.}}
\maketitle
%
%\author{Lei Yu\thanks{L. Yu is with the School of Statistics and Data Science, LPMC, KLMDASR,
%and LEBPS, Nankai University, Tianjin 300071, China (e-mail: leiyu@nankai.edu.cn).
%This work was supported by the NSFC under grant 62101286 and the Fundamental
%Research Funds for the Central Universities of China (Nankai University)
%under grant 054-63233073.}}
%\maketitle
\begin{abstract}
Nonconvex optimization plays a key role in multi-user information
theory and related fields, but it is usually difficult to solve. The
rate region of the Gray--Wyner  source coding system (or almost
equivalently, the mutual information region) is a typical example
in nonconvex optimization, whose single-letter expression was given
by Gray and Wyner. However, due to the nonconvexity of the optimization
involved in this expression, previously, there was none nontrivial
discrete source for which the analytic expression is known. In this
paper, we propose a new strategy to solve nonconvex optimization problems.
By this strategy, we provide the analytic expression for the doubly
symmetric binary source (DSBS), which confirms positively a conjecture
of Gray and Wyner in 1974.  We also provide the analytic expression
of the mutual information region for the Gaussian source, and provide
(or recover) the analytic expressions of  the lossy Gray--Wyner
region for both the DSBS and Gaussian source. Our proof strategy relies
on  an auxiliary measure technique and the analytical expression
of the optimal-transport divergence region. 

\end{abstract}


\section{Introduction}

\begin{figure}
\centering 

\tikzset{every picture/.style={line width=0.75pt}}   %set default line width to 0.75pt        
\begin{tikzpicture}[x=0.75pt,y=0.75pt,yscale=-1,xscale=1,scale=0.8]
 %uncomment if require: \path (0,300);   %set diagram left start at 0, and has height of 300
 %Rounded Rect [id:dp7300319029719384] 
\draw   (159.8,118.8) .. controls (159.8,114.38) and (163.38,110.8) .. (167.8,110.8) -- (221.8,110.8) .. controls (226.22,110.8) and (229.8,114.38) .. (229.8,118.8) -- (229.8,142.8) .. controls (229.8,147.22) and (226.22,150.8) .. (221.8,150.8) -- (167.8,150.8) .. controls (163.38,150.8) and (159.8,147.22) .. (159.8,142.8) -- cycle ;
 %Rounded Rect [id:dp10758393863407445] 
\draw   (310.2,45.2) .. controls (310.2,40.78) and (313.78,37.2) .. (318.2,37.2) -- (382.52,37.2) .. controls (386.94,37.2) and (390.52,40.78) .. (390.52,45.2) -- (390.52,69.2) .. controls (390.52,73.62) and (386.94,77.2) .. (382.52,77.2) -- (318.2,77.2) .. controls (313.78,77.2) and (310.2,73.62) .. (310.2,69.2) -- cycle ;
 %Rounded Rect [id:dp4641924753255342] 
\draw   (309.4,189.2) .. controls (309.4,184.78) and (312.98,181.2) .. (317.4,181.2) -- (383.32,181.2) .. controls (387.74,181.2) and (391.32,184.78) .. (391.32,189.2) -- (391.32,213.2) .. controls (391.32,217.62) and (387.74,221.2) .. (383.32,221.2) -- (317.4,221.2) .. controls (312.98,221.2) and (309.4,217.62) .. (309.4,213.2) -- cycle ;
 %Straight Lines [id:da554133192319858] 
\draw    (196.92,151.16) -- (196.92,201.56) -- (307.72,201.56) ;
\draw [shift={(309.72,201.56)}, rotate = 180] [color={rgb, 255:red, 0; green, 0; blue, 0 }  ][line width=0.75]    (10.93,-3.29) .. controls (6.95,-1.4) and (3.31,-0.3) .. (0,0) .. controls (3.31,0.3) and (6.95,1.4) .. (10.93,3.29)   ;
 %Straight Lines [id:da7347006786308408] 
\draw    (196.12,110.36) -- (196.12,59.16) -- (307.72,59.16) ;
\draw [shift={(309.72,59.16)}, rotate = 180] [color={rgb, 255:red, 0; green, 0; blue, 0 }  ][line width=0.75]    (10.93,-3.29) .. controls (6.95,-1.4) and (3.31,-0.3) .. (0,0) .. controls (3.31,0.3) and (6.95,1.4) .. (10.93,3.29)   ;
 %Straight Lines [id:da25444122368949196] 
\draw    (229.72,130.2) -- (352.92,130.2) -- (352.92,79.56) ;
\draw [shift={(352.92,77.56)}, rotate = 90] [color={rgb, 255:red, 0; green, 0; blue, 0 }  ][line width=0.75]    (10.93,-3.29) .. controls (6.95,-1.4) and (3.31,-0.3) .. (0,0) .. controls (3.31,0.3) and (6.95,1.4) .. (10.93,3.29)   ;
 %Straight Lines [id:da33990650977837733] 
\draw    (229.72,130.2) -- (352.92,130.2) -- (352.92,177.96) ;
\draw [shift={(352.92,179.96)}, rotate = 270] [color={rgb, 255:red, 0; green, 0; blue, 0 }  ][line width=0.75]    (10.93,-3.29) .. controls (6.95,-1.4) and (3.31,-0.3) .. (0,0) .. controls (3.31,0.3) and (6.95,1.4) .. (10.93,3.29)   ;
 %Straight Lines [id:da3951763901667362] 
\draw    (78.52,129.56) -- (158.12,129.56) ;
\draw [shift={(160.12,129.56)}, rotate = 180] [color={rgb, 255:red, 0; green, 0; blue, 0 }  ][line width=0.75]    (10.93,-3.29) .. controls (6.95,-1.4) and (3.31,-0.3) .. (0,0) .. controls (3.31,0.3) and (6.95,1.4) .. (10.93,3.29)   ;
 %Straight Lines [id:da5910425784133002] 
\draw    (390.52,57.56) -- (458.92,57.56) ;
\draw [shift={(460.92,57.56)}, rotate = 180] [color={rgb, 255:red, 0; green, 0; blue, 0 }  ][line width=0.75]    (10.93,-3.29) .. controls (6.95,-1.4) and (3.31,-0.3) .. (0,0) .. controls (3.31,0.3) and (6.95,1.4) .. (10.93,3.29)   ;
 %Straight Lines [id:da11477359446957047] 
\draw    (391.32,199.96) -- (459.72,199.96) ;
\draw [shift={(461.72,199.96)}, rotate = 180] [color={rgb, 255:red, 0; green, 0; blue, 0 }  ][line width=0.75]    (10.93,-3.29) .. controls (6.95,-1.4) and (3.31,-0.3) .. (0,0) .. controls (3.31,0.3) and (6.95,1.4) .. (10.93,3.29)   ;
 % Text Node
\draw (164.6,122) node [anchor=north west][inner sep=0.75pt]   [align=left] {Encoder};
 % Text Node
\draw (315,49) node [anchor=north west][inner sep=0.75pt]   [align=left] {Decoder 1};
 % Text Node
\draw (314.2,193) node [anchor=north west][inner sep=0.75pt]   [align=left] {Decoder 2};
 % Text Node
\draw (256.4,107) node [anchor=north west][inner sep=0.75pt]    {$R_{0}$};
 % Text Node
\draw (256.4,35) node [anchor=north west][inner sep=0.75pt]    {$R_{1}$};
 % Text Node
\draw (257.2,177.4) node [anchor=north west][inner sep=0.75pt]    {$R_{2}$};
 % Text Node
\draw (76.8,98.4) node [anchor=north west][inner sep=0.75pt]    {$\left( X^{n} ,Y^{n}\right)$};
 % Text Node
\draw (412.92,32.4) node [anchor=north west][inner sep=0.75pt]    {$X^{n}$};
 % Text Node
\draw (413.72,174.8) node [anchor=north west][inner sep=0.75pt]    {$Y^{n}$};
\end{tikzpicture} \caption{\label{fig:Gray=002013Wyner-coding-system.}Gray--Wyner coding system.}

\end{figure}
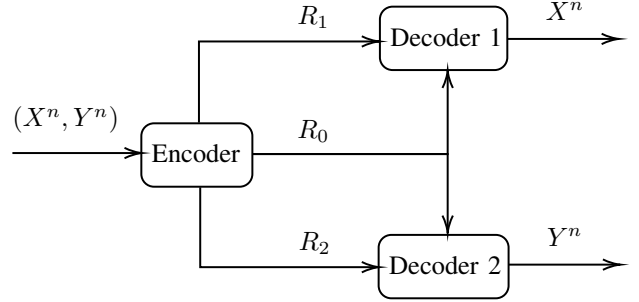

The Gray--Wyner coding system illustrated in Fig. \ref{fig:Gray=002013Wyner-coding-system.}
was initially investigated by Gray and Wyner in a seminal work \cite{GrayWyner},
and then widely investigated in the literature; see, e.g., \cite{kamath2010new,watanabe2016second,zhou2016discrete,li2017extended,graczyk2020gray,sula2021gray,Visw}.
In this system, two correlated memoryless sources $X^{n},Y^{n}$ are
respectively required to be transmitted almost losslessly from one
sender to two receivers. The joint distribution of these sources is
denoted by $P_{XY}$ which is assumed to be defined on finite alphabets.
Both the decoders are connected to the encoder by a common channel,
and each decoder is also connected to the encoder by its own private
channel. All these channels are noiseless. The common rate is denoted
by $R_{0}$ and the private rates are respectively denoted by $R_{1}$
and $R_{2}$. The (lossless) Gray--Wyner rate region is the set of
$(R_{0},R_{1},R_{2})$ such that the sources $X^{n},Y^{n}$ can be
transmitted almost losslessly by using some code with rates $(R_{0},R_{1},R_{2})$.
Gray and Wyner showed that the Gray--Wyner rate region is equal to
the set 
\begin{align*}
\mathcal{R} & :=\bigl\{(R_{0},R_{1},R_{2}):\exists P_{W|XY},R_{0}\ge I(X,Y;W),\\
 & \qquad\qquad\qquad R_{1}\ge H(X|W),R_{2}\ge H(Y|W)\bigr\}.
\end{align*}
The cardinality of the alphabet of $W$ can be assumed no larger than
$\left|\mathcal{X}\right|\left|\mathcal{Y}\right|+2$. This is a single-letter
characterization which means the expression is independent of the
dimension (or blocklength). The region $\mathcal{R}$ is obviously
determined by its lower envelope which is given by
\begin{align}
R_{0}(R_{1},R_{2}) & :=\inf\left\{ R_{0}:(R_{0},R_{1},R_{2})\in\mathcal{R}\right\} \nonumber \\
 & =\inf_{\substack{P_{W|XY}:H(X|W)\le R_{1},\\
H(Y|W)\le R_{2}
}
}I(X,Y;W).\label{eq:R0}
\end{align}

Solving this optimization is in fact a   difficult open question,
due to its nonconvexity. In fact, previously, there was even none
nontrivial case for which the analytic expression is known.  Gray
and Wyner \cite{GrayWyner} tried to provide an analytic expression
for the doubly symmetric binary source (DSBS), and made a conjecture.
Consider a DSBS with disagree probability $p\in(0,1/2)$, whose distribution,
denoted by $\ensuremath{\DSBS(p)}$, is given in Table \ref{tab:The-distribution-of}.
 In other words, for $(X,Y)\sim\ensuremath{\DSBS(p)}$, $X$ is a
Bernoulli random variable with parameter $1/2$, and $Y$ is the output
distribution of a binary symmetric channel $\BSC(p)$ with crossover
probability $p$ when the input is $X$. For such a DSBS, its rate
distortion function under the Hamming distortion $d_{\mathrm{H}}$
is given by \cite{GrayWyner,nayak2010successive}
\begin{align}
 & R(D_{1},D_{2})\nonumber \\
 & :=\inf_{\substack{P_{UV|XY}:\mathbb{E}d_{\mathrm{H}}(X,U)\le D_{1},\\
\mathbb{E}d_{\mathrm{H}}(Y,U)\le D_{2}
}
}I(X,Y;U,V)\nonumber \\
 & =\begin{cases}
1-(1-p)h\left(\frac{a+b-p}{2(1-p)}\right)\\
\qquad\qquad-ph\left(\frac{a-b+p}{2p}\right), & a*p\ge b,a*b\ge p\\
1+h(p)-h(a)-h(b), & a*b\le p\\
1-h(a), & a*p\le b
\end{cases}\label{eq:-18}
\end{align}
with $b=D_{1}\lor D_{2},a=D_{1}\land D_{2}$, where $h:t\mapsto-t\log t-(1-t)\log(1-t)$
denotes the binary entropy function, and $h^{-1}$ is the inverse
of the restriction of $h$ to the set $\left[0,\frac{1}{2}\right]$.
Here, $x\lor y:=\max\{x,y\}$, $x\land y=\min\{x,y\}$, and $a*b=a(1-b)+b(1-a)$
is the binary convolution operation. Throughout this paper, for the
DSBS, we always use the logarithm with base $2$, denoted by $\log$,
and for Gaussian sources, always use the one with natural base, denoted
by $\ln$. For the DSBS, Gray and Wyner \cite{GrayWyner} made the
following conjecture.

\begin{table} 
\begin{centering}  
\caption{\label{tab:The-distribution-of} The distribution of a DSBS with parameter $p\in (0,1/2)$, which is denoted by $\DSBS(p)$.}
\begin{tabular}{|p{1cm}|p{1cm}|p{1cm}|} \hline  $$X\backslash Y$$ & $$0$$ & $$1$$\tabularnewline \hline  $$0$$ &  $$\frac{1-p}{2}$$ & $$\frac{p}{2}$$\tabularnewline \hline  $$1$$ & $$\frac{p}{2}$$ & $$\frac{1-p}{2}$$\tabularnewline \hline  \end{tabular} \par 
\end{centering}  \end{table}
\begin{conjecture}
\label{conj:-For-the}\cite{GrayWyner,wyner1974recent} For the source
$\ensuremath{\DSBS(p)}$,  it holds that for $(R_{1},R_{2})\in[0,1]^{2}$,
\[
R_{0}(R_{1},R_{2})=R(h^{-1}(R_{1}),h^{-1}(R_{2})).
\]
\end{conjecture}
This conjecture has been open for nearly 50 years since 1974. Although
there are now a vast number of works existing in the literature 
on the Gray--Wyner coding system, surprisingly, there seems no progress
on this conjecture until now. The intuition behind this conjecture
is that the sender first encodes the source into $W=(U,V)$ by using
an optimal point-to-point lossy compression code with distortions
$(D_{1},D_{2})$ and rate $R(D_{1},D_{2})$, and then compress $X\oplus U$
and $Y\oplus V$ losslessly using rate $h(D_{1})$ and $h(D_{2})$
respectively. Here we choose $(D_{1},D_{2})=(h^{-1}(R_{1}),h^{-1}(R_{2}))$.
In other words, the Gray--Wyner conjecture above states that this
layered coding scheme is optimal for the Gray--Wyner system for the
DSBS. 

The Gray--Wyner region can be also expressed by the mutual information
region. Given an arbitrary (not necessarily discrete) joint distribution
$P_{XY}$, define the mutual information region as 
\begin{align*}
\mathcal{I} & :=\mathcal{I}(P_{XY})\\
 & :=\left\{ \left(I(X;W),I(Y;W),I(X,Y;W)\right)\right\} _{P_{W|XY}}.
\end{align*}
Its projection region on the plane of the first two coordinates is
\[
\mathcal{I}_{0}:=\left\{ \left(I(X;W),I(Y;W)\right)\right\} _{P_{W|XY}}.
\]
The mutual information region is determined by its lower and upper
envelopes which are respectively defined as for $(\alpha,\beta)\in\mathcal{I}_{0}$,
\begin{align*}
\underline{\Upsilon}(\alpha,\beta) & :=\inf\left\{ \gamma:(\alpha,\beta,\gamma)\in\mathcal{I}\right\} \\
 & =\inf_{P_{W|XY}:I(X;W)=\alpha,I(Y;W)=\beta}I(X,Y;W),
\end{align*}
and 
\begin{align*}
\overline{\Upsilon}(\alpha,\beta) & :=\sup\left\{ \gamma:(\alpha,\beta,\gamma)\in\mathcal{I}\right\} \\
 & =\sup_{P_{W|XY}:I(X;W)=\alpha,I(Y;W)=\beta}I(X,Y;W).
\end{align*}
We also define the lower increasing envelope as 
\begin{align}
\Upsilon(\alpha,\beta) & :=\inf\left\{ \gamma:(\alpha,\beta,\gamma)\in\mathcal{I}\right\} \nonumber \\
 & =\inf_{P_{W|XY}:I(X;W)\ge\alpha,I(Y;W)\ge\beta}I(X,Y;W).\label{eq:mi-4}
\end{align}

Observe that $R_{0}(R_{1},R_{2})=\Upsilon(H(X)-R_{1},H(Y)-R_{2})$,
and hence, characterizing $R_{0}(R_{1},R_{2})$ is equivalent to characterizing
$\Upsilon(\alpha,\beta)$. In this paper we only focus on $\Upsilon(\alpha,\beta)$
and also $\underline{\Upsilon}(\alpha,\beta)$ and $\overline{\Upsilon}(\alpha,\beta)$.
In fact, the function $\Upsilon(\alpha,\beta)$ is determined by $\underline{\Upsilon}(\alpha,\beta)$.
It should be also noted that the function $\Upsilon(\alpha,\beta)$
for Gaussian distributions was already expressed in terms of optimizations
over Gaussian random variables by using the doubling trick \cite{nair2014extremal,liu2018information}.
 Furthermore, the mutual information region can be also expressed
in terms of the conditional entropy region $\mathcal{H}:=\left\{ \left(H(X|W),H(Y|W),H(X,Y|W)\right)\right\} _{P_{W|XY}}.$ 

\subsection{\label{subsec:Our-Contributions}Our Contributions }

The main difficulty in proving Conjecture \ref{conj:-For-the} is
that the optimization problem  in \eqref{eq:R0} (or \eqref{eq:mi-4})
is nonconvex. One routine strategy to solve nonconvex optimization
is to apply Karush--Kuhn--Tucker (KKT) conditions to obtain several
necessary optimality equations, and then solve these equations to
find the optimal solution. However, it is a challenge to solve these
optimality equations in this setting, since logarithmic functions
are involved in them. Therefore, new techniques are required to resolve
Conjecture \ref{conj:-For-the}. 

In fact,  nonconvex optimization is very common in today's information
theory, which originated with accompanied by multi-user information
theory; e.g., Mrs. Gerber's lemma \cite{wyner1973theorem,witsenhausen1974entropy,tishby2000information}
and Wyner's common information \cite{WynerCI} which both involves
nonconvex optimization. Although this kind of problems exist in the
literature for a long time, nowadays relatively little is known about
them. In other words, finding new ideas to solve nonconvex optimization
is a very difficult task. During the last decade, Nair and his collaborators
have made some significant contributions in this field; see e.g.,
\cite{nair2014extremal,geng2014capacity,ding2021concavity,lau2022uniqueness}.
For example, the change of variables technique was exploited by them
to convert a nonconvex optimization problem to a convex one. However,
such a technique seems failed to be applied directly to the optimization
problem  in \eqref{eq:R0} (or \eqref{eq:mi-4}). Readers can refer
to \cite{wang2021optimization} for recent advances in this field,
especially for optimizations for discrete distributions. 

In this paper, we propose a new strategy to solve nonconvex optimization
problems. By this strategy, we confirm Conjecture \ref{conj:-For-the}
positively, which yields the first explicit expression for the Gray--Wyner
region of a certain source. We also prove the analytic expression
of the mutual information region for the Gaussian source, and also
prove (or recover) the analytic expressions of  the lossy Gray--Wyner
region for both the DSBS and Gaussian source.  Our proof strategy
integrates  an auxiliary measure technique with the convexity of
the  envelopes of the optimal-transport divergence region \cite{yu2021convexity_article}
for the DSBS, and integrates the same auxiliary measure technique
with hypercontractivity inequalities for the Gaussian source. For
the Gaussian Gray--Wyner system, the lossy Gray--Wyner region was
previously presented in \cite{chen2021computing} and partially in
\cite{sula2021gray} by using methods different from ours.  It
is worth noting that our convexity result derived in \cite{yu2021convexity_article}
turns out to be important, since it is not only used in \cite{yu2021convexity_article}
as a key ingredient in the proof of the Ordentlich--Polyanskiy--Shayevitz
conjecture \cite{ordentlich2020note} (which is a conjecture on the
strong version of the small-set expansion theorem), but also used
in this paper to resolve the Gray--Wyner conjecture.  Both the auxiliary
measure technique and the convexity result in \cite{yu2021convexity_article}
are indispensable in our proofs, which makes our proofs nontrivial. 

\subsection{Notations}

We use $X\stackrel{P_{Y|X}}{\longrightarrow}Y$ to denote that the
random variable $Y$ is the output of the channel $P_{Y|X}$ when
the input is $X$. We denote $\BSC(a)$ as a binary symmetric channel
with crossover probability $a$. We denote $\DSBS(p)$ as the DSBS
with disagree probability $p$, and $\mathrm{Bern}(a)$ as the Bernoulli
distribution with parameter $a$. For a real-valued function $f,$
we denote $\conv f$ and $\conc f$ respectively as the lower convex
envelope and the upper concave envelope of $f$. 

\textbf{}

\section{Main Results}

\subsection{Mutual Information Region for DSBS }

 In this subsection and in the corresponding proofs of results stated
in this subsection, we use the logarithm with base $2$, which is
denoted by $\log$.

\begin{figure*}
\centering%
\begin{tabular}{cc}
\includegraphics[width=0.45\textwidth]{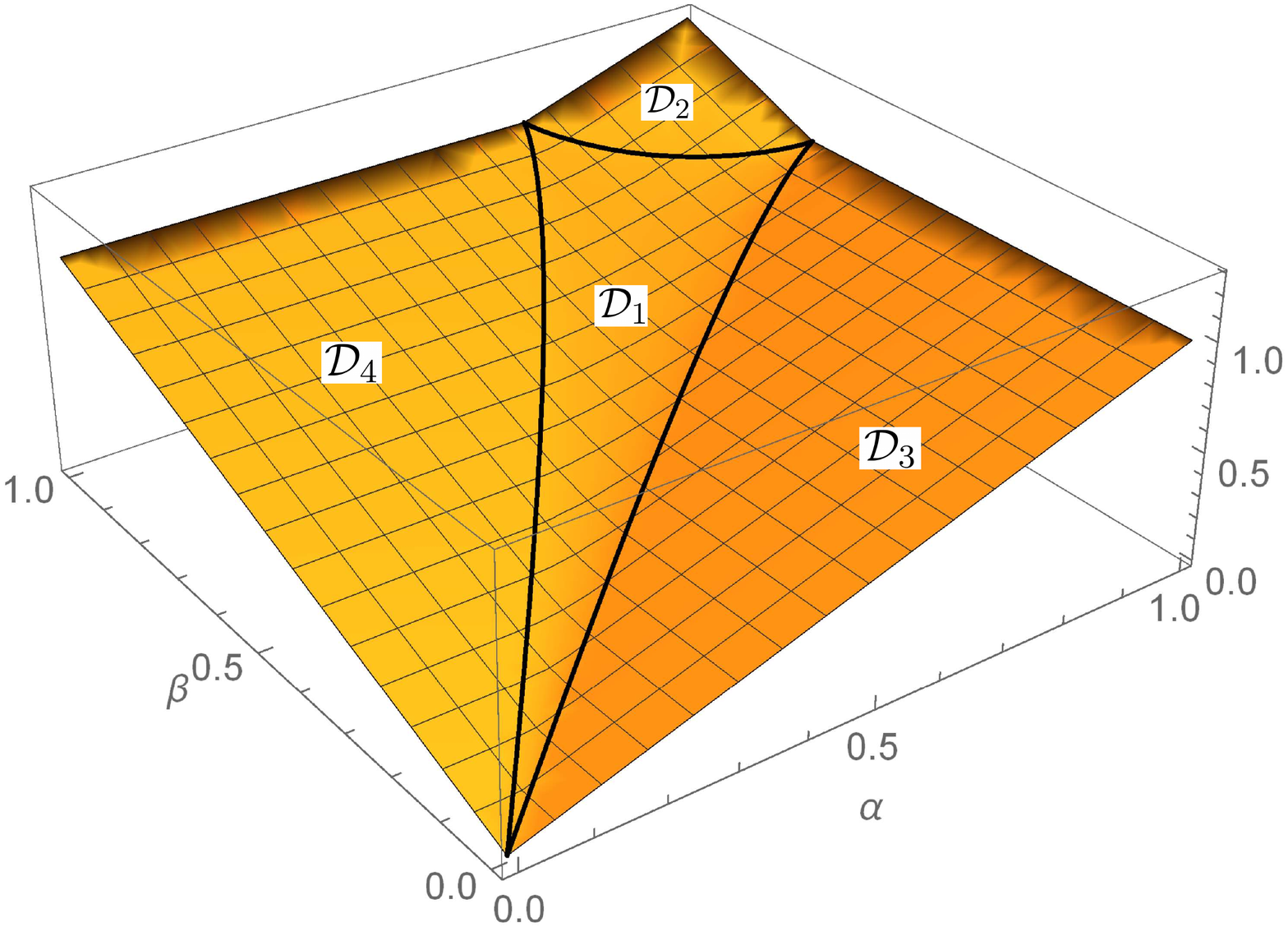} & \includegraphics[width=0.45\textwidth]{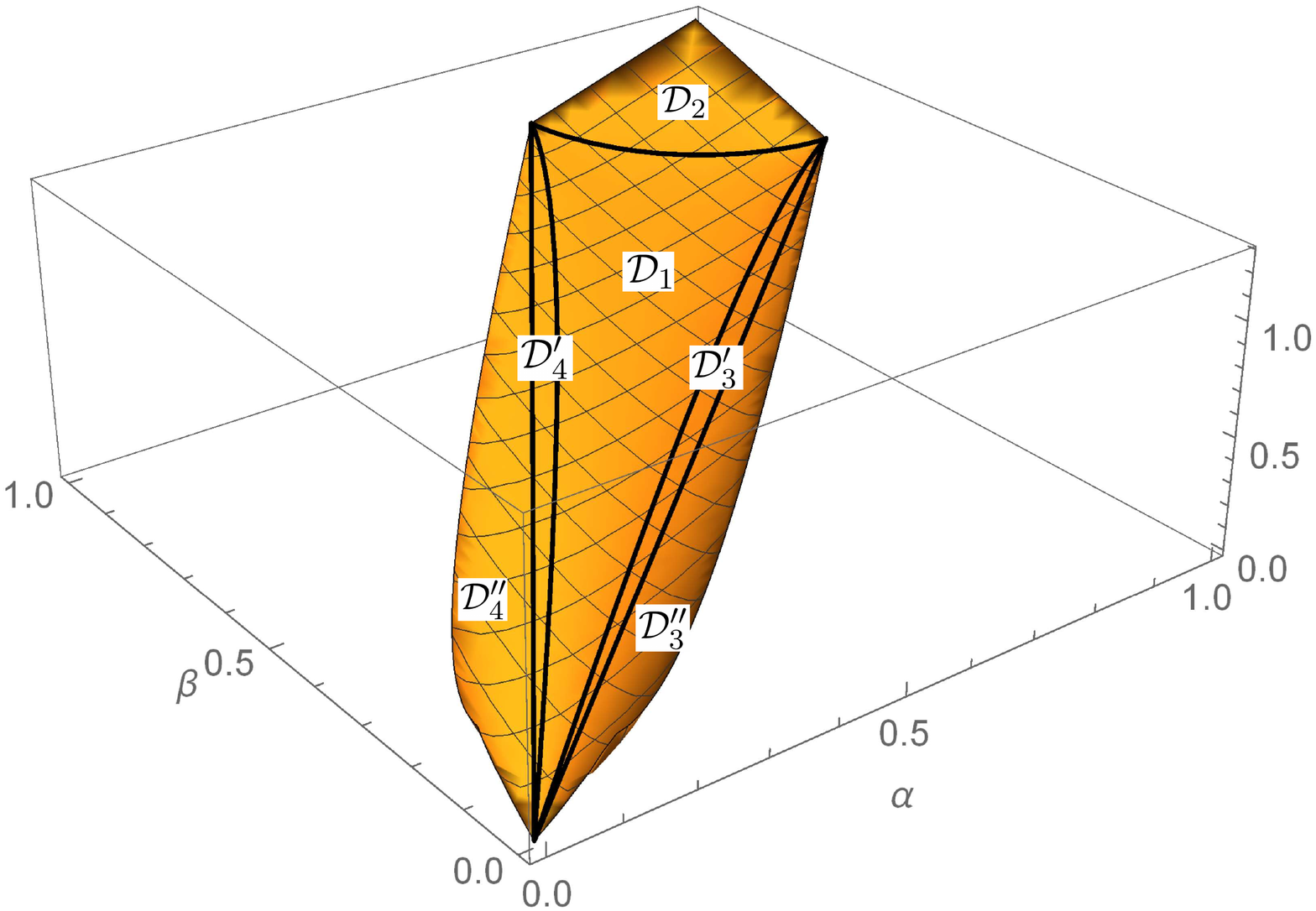}\tabularnewline
{\footnotesize{}(a) $\Upsilon^{*}$} & {\footnotesize{}(b) $\underline{\Upsilon}^{*}$}\tabularnewline
\end{tabular}

\begin{tabular}{c}
\includegraphics[width=0.45\textwidth]{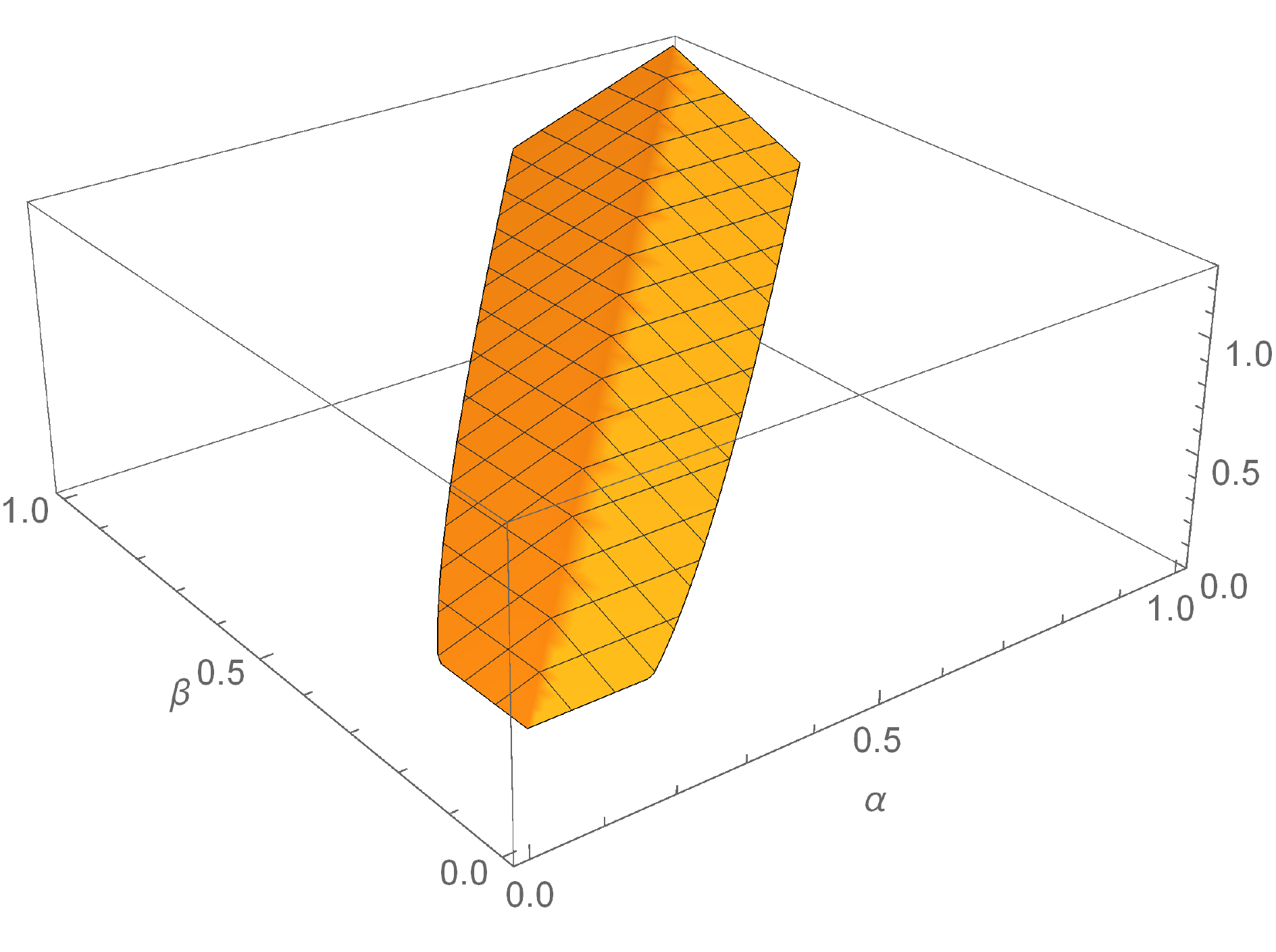}\tabularnewline
{\footnotesize{}(c) $\overline{\Upsilon}^{*}$}\tabularnewline
\end{tabular} 

\caption{\label{fig:upsilon}Illustration of $\Upsilon^{*}$, $\underline{\Upsilon}^{*}$,
and $\overline{\Upsilon}^{*}$ for $p=0.05$ (equivalently, the correlation
coefficient $\rho=0.9$). The boundaries of the graphs of  $\underline{\Upsilon}^{*}$
and $\overline{\Upsilon}^{*}$ coincide except at $(\alpha,\beta)$
belonging to a neighborhood of the origin. }
\end{figure*}

 For $(\alpha,\beta)\in[0,1]^{2}$, denote $a=h^{-1}(1-\alpha),b=h^{-1}(1-\beta)$.
Define several disjoint sets 
\begin{align*}
\mathcal{D}_{1} & :=\{(\alpha,\beta)\in[0,1]^{2}:a*p\ge b,\;\\
 & \qquad\qquad b*p\ge a,\;a*b\ge p\},\\
\mathcal{D}_{2} & :=\{(\alpha,\beta)\in[0,1]^{2}:a*b<p\},\\
\mathcal{D}_{3} & :=\{(\alpha,\beta)\in[0,1]^{2}:a*p<b\},\\
\mathcal{D}_{4} & :=\{(\alpha,\beta)\in[0,1]^{2}:b*p<a\}.
\end{align*}
For $(\alpha,\beta)\in[0,1]^{2}$, define 
\begin{equation}
\Upsilon^{*}(\alpha,\beta):=\begin{cases}
1-(1-p)h\left(\frac{a+b-p}{2(1-p)}\right)\\
\qquad\qquad-ph\left(\frac{a-b+p}{2p}\right), & (\alpha,\beta)\in\mathcal{D}_{1}\\
1+h(p)-h(a)-h(b), & (\alpha,\beta)\in\mathcal{D}_{2}\\
1-h(a), & (\alpha,\beta)\in\mathcal{D}_{3}\\
1-h(b), & (\alpha,\beta)\in\mathcal{D}_{4}
\end{cases}.\label{eq:mi}
\end{equation}
In fact, $\Upsilon^{*}(\alpha,\beta)$ is nothing but $R(h^{-1}(1-\alpha),h^{-1}(1-\beta))$,
where $R(\cdot,\cdot)$ is the rate-distortion function for $\ensuremath{\DSBS(p)}$
given in \eqref{eq:-18}. The following is one of our main results,
whose proof is given in Section \ref{sec:Proof-of-Theorem-mi}.
\begin{thm}[Gray--Wyner Region for DSBS]
 \label{thm:mi} For the source $\ensuremath{\DSBS(p)}$ with $p\in(0,1/2)$,
it holds that for $(\alpha,\beta)\in[0,1]^{2}$, 
\begin{align*}
\Upsilon(\alpha,\beta) & =\Upsilon^{*}(\alpha,\beta).
\end{align*}
\end{thm}
Observe that by definitions, $R_{0}(R_{1},R_{2})=\Upsilon(1-R_{1},1-R_{2})$
and $\Upsilon^{*}(1-R_{1},1-R_{2})=R(h^{-1}(R_{1}),h^{-1}(R_{2}))$,
 where $R_{0}(\cdot,\cdot)$ and $R(\cdot,\cdot)$ are respectively
 given in \eqref{eq:R0} and \eqref{eq:-18}. So, Theorem \ref{thm:mi}
implies $R_{0}(R_{1},R_{2})=R(h^{-1}(R_{1}),h^{-1}(R_{2}))$, which
confirms Conjecture \ref{conj:-For-the} positively. The function
$\Upsilon^{*}$ is plotted in Fig. \ref{fig:upsilon}. 

As mentioned in Section \ref{subsec:Our-Contributions}, the main
difficulty to prove Theorem \ref{thm:mi} is the nonconvexity of the
optimization involved in the definition of $\Upsilon$ (see \eqref{eq:mi-4}).
One might plan to use Karush--Kuhn--Tucker (KKT) conditions to obtain
several necessary optimality equations, and then solve these equations
to find the optimal solution. However, solving these equations is
a challenge, due to the fact that logarithmic functions are involved.
Instead, we propose the following strategy to prove Theorem \ref{thm:mi},
which consists of two steps.
\begin{enumerate}
\item Note that the optimization in \eqref{eq:mi-4} can be written as the
one over $Q_{WXY}$ with the marginal constraint $Q_{XY}=P_{XY}$.
In this step, we introduce an auxiliary probability measure $R_{XY}$,
and by the formula $D(Q_{Z|W}\|R_{Z}|Q_{W})-D(P_{Z}\|R_{Z})$ with
$Z=X,$ $Y$, or $(X,Y)$, rewrite all the mutual informations in
the objective function or the constraints as relative entropies (since
the latter are easier to deal with). Then, relax the optimization
problem by discarding the marginal constraint $Q_{XY}=P_{XY}$. That
is, we obtain a new optimization problem which only involves relative
entropies (with fixed distribution $R$ as the second arguments).
\item The new optimization problem obtained above is in fact an optimization
over the time-sharing variable (or convex-combination variable) $W$.
In other words, the value of this new optimization problem is determined
by the lower convex envelope of the relative entropy region $\{(D(Q_{X}\|R_{X}),D(Q_{Y}\|R_{Y}),D(Q_{XY}\|R_{XY}))\}_{Q_{XY}}$.
Hence, to solve this new optimization problem, more specifically,
to remove the time-sharing variables, it suffices to prove the convexity
of this lower convex envelope.  This part has been done in our another
work \cite{yu2021convexity_article}, or see Lemma \ref{lem:convexity}
in Section \ref{sec:Preliminaries-on-Optimal-Transpo}. The proof
therein relies on a new technique, called the first-order method,
which is based on the equivalence between the convexity of a function
and the convexity of the set of minimizers of its Lagrangian dual.
 Denote the optimal solution to the new optimization by $Q_{XY}^{*}$.
\end{enumerate}
In Step 1, to make the optimization problem simpler, we would like
to discard the marginal constraint $Q_{XY}=P_{XY}$. Although we can
discard it directly without introducing the auxiliary measure $R_{XY}$,
the resultant bound would be far from optimal. In other words, the
role of the auxiliary measure is that by properly choosing this measure,
it enables us not to lose too much when we discard the marginal constraint.
To ensure that the bound derived by the method above is tight, we
need choose the $R_{XY}$ as an optimal distribution (called shadow
measure), which can be specified in the following way. 

For the DSBS, denote $P_{W|XY}^{*}$ as an optimal distribution attaining
the infimum in \eqref{eq:mi-4} (i.e., the one in the Gray--Wyner
conjecture). In fact, the distribution $P_{XY|W}^{*}$ induced by
$P_{WXY}^{*}:=P_{W|XY}^{*}P_{XY}$ satisfies certain symmetry so that
given any DSBS $S_{XY}$, $D(P_{XY|W}^{*}\|S_{XY}|P_{W}^{*})=D(P_{XY|W=w}^{*}\|S_{XY})$
holds for any $w$. In our proof, we choose $R_{XY}$ as a  DSBS
for which the optimal solution $Q_{XY}^{*}$ in Step 2 is exactly
$P_{XY|W=w}^{*}$ for some $w$. Hence, the final bound obtained in
Step 2 is 
\begin{align*}
 & D(Q_{XY}^{*}\|R_{XY})-D(P_{XY}\|R_{XY})\\
 & =D(P_{XY|W=w}^{*}\|R_{XY})-D(P_{XY}\|R_{XY})\\
 & =D(P_{XY|W}^{*}\|R_{XY}|P_{W}^{*})-D(P_{XY}\|R_{XY})\\
 & =I_{P^{*}}(X,Y;W),
\end{align*}
where the last line follows since the $(X,Y)$-marginal of $P_{WXY}^{*}$
is exactly $P_{XY}$. Therefore, the bound induced by such $R_{XY}$
is tight. In other words, such a choice of $R_{XY}$ is optimal. 

As mentioned in Section \ref{subsec:Our-Contributions}, our convexity
result in \cite{yu2021convexity_article} was previously used as a
key ingredient in the proof of the Ordentlich--Polyanskiy--Shayevitz
conjecture \cite{ordentlich2020note}; refer to \cite{yu2021convexity_article}
for more details. Interestingly, it also can be used as a key tool
to resolve the Gray--Wyner conjecture (in Step 2) in this paper.
This forces us to re-examine the importance of the convexity result
in \cite{yu2021convexity_article}. At the technical level, our proof
strategy in present paper integrates two techniques: the auxiliary
measure method (in Step 1) and the first-order method (in Step 2).
These two indispensable techniques are nontrivial on their own, which
hence in turn makes our proof nontrivial.  Furthermore, although
we only consider the optimization with marginal distributions fixed,
we believe that our strategy above can be also applied to many other
similar optimization problems, e.g.,  optimizations in which a channel
is fixed and the input of this channel is  to be optimized. 

As a consequence of Theorem \ref{thm:mi}, the rate-distortion region
of the lossy Gray--Wyner system can be obtained. In the Gray--Wyner
system, consider a distortion measure $d$. If the reconstructions
of the sources at two receivers are allowed to be within distortion
levels $D_{1}$ and $D_{2}$ respectively, then the rate-distortion
region is defined as the set of tuples $(R_{0},R_{1},R_{2},D_{1},D_{2})$.
Such a region was shown by Gray and Wyner \cite{GrayWyner} to be
\begin{align*}
\mathcal{R}_{\mathrm{lossy}}:= & \bigl\{(R_{0},R_{1},R_{2},D_{1},D_{2}):\\
 & \qquad\exists P_{W|XY},P_{\hat{X}|WX},P_{\hat{Y}|WY},\\
 & \qquad R_{0}\ge I(X,Y;W),\\
 & \qquad R_{1}\ge I(X;\hat{X}|W),R_{2}\ge I(Y;\hat{Y}|W),\\
 & \qquad\mathbb{E}d(X,\hat{X})\le D_{1},\mathbb{E}d(Y,\hat{Y})\le D_{2}\bigr\}.
\end{align*}
Computing this region is equivalent to computing the following function
\begin{align}
 & R_{0}(R_{1},R_{2},D_{1},D_{2})\nonumber \\
 & :=\inf_{\substack{P_{W|XY},P_{\hat{X}|WX},P_{\hat{Y}|WY}:\\
I(X;\hat{X}|W)\le R_{1},I(Y;\hat{Y}|W)\le R_{2},\\
\mathbb{E}d(X,\hat{X})\le D_{1},\mathbb{E}d(Y,\hat{Y})\le D_{2}
}
}I(X,Y;W).\label{eq:lossyRD}
\end{align}
Using Theorem \ref{thm:mi}, we obtain the analytical expression for
this function. 
\begin{cor}[Lossy Gray--Wyner Rate Region for DSBS]
\label{cor:lossyGW}  For the source $\ensuremath{\DSBS(p)}$ with
$p\in(0,1/2)$, under the Hamming distortion measure,  it holds
that for $R_{1},R_{2},D_{1},D_{2}\ge0$, 
\begin{align*}
 & R_{0}(R_{1},R_{2},D_{1},D_{2})\\
 & =\Upsilon^{*}(\left[R(D_{1})-R_{1}\right]^{+},\left[R(D_{2})-R_{2}\right]^{+}),
\end{align*}
where $\Upsilon^{*}$ is defined in \eqref{eq:mi}, $\left[x\right]^{+}:=\max\{x,0\}$,
and $R(D):=1-h(D)$ is the rate-distribution function of Bernoulli
source $\mathrm{Bern}(\frac{1}{2})$. 
\end{cor}
\begin{IEEEproof}
For a feasible tuple $(P_{W|XY},P_{\hat{X}|WX},P_{\hat{Y}|WY})$ satisfying
the constraints in \eqref{eq:lossyRD},  it holds that 
\begin{align*}
I(X;W) & \ge\left[I(X;\hat{X}W)-R_{1}\right]^{+}\\
 & \ge\left[I(X;\hat{X})-R_{1}\right]^{+}\\
 & \ge\left[R(D_{1})-R_{1}\right]^{+},
\end{align*}
and similarly,
\begin{align*}
I(Y;W) & \ge\left[R(D_{2})-R_{2}\right]^{+}.
\end{align*}
Therefore, 
\begin{align*}
 & R_{0}(R_{1},R_{2},D_{1},D_{2})\\
 & \ge\inf_{\substack{P_{W|XY}:I(X;W)\ge\left[R(D_{1})-R_{1}\right]^{+}\\
I(Y;W)\ge\left[R(D_{2})-R_{2}\right]^{+}
}
}I(X,Y;W)\\
 & =\Upsilon(\left[R(D_{1})-R_{1}\right]^{+},\left[R(D_{2})-R_{2}\right]^{+}).
\end{align*}

We now prove the other direction. From the proof of Theorem \ref{thm:mi},
there is a conditional distribution $P_{W|XY}$ attaining $\Upsilon(\left[R(D_{1})-R_{1}\right]^{+},\left[R(D_{2})-R_{2}\right]^{+})$
(i.e., the infimum in \eqref{eq:mi-4})  such that\footnote{Rigorously speaking, $(W,X)$ and $(W,Y)$ are not always DSBSes,
since in some case of our proof, $W=(U,V)$ such that $X\stackrel{\BSC(a)}{\longrightarrow}U\stackrel{\BSC(c)}{\longrightarrow}V\stackrel{\BSC(b)}{\longrightarrow}Y$,
and hence $(U,X)$ and $(V,Y)$ are DSBSes. However, for this case,
the argument given here with slight modification still works.} both $(W,X)$ and $(W,Y)$ are DSBSes.  It is well known that for
a DSBS $(W,X)$, we can write $X\stackrel{\BSC(\theta_{1})}{\longrightarrow}\hat{X}\stackrel{\BSC(\theta_{2})}{\longrightarrow}W$
for any parameters $\theta_{1},\theta_{2}\in[0,1]$ such that $\theta_{1}*\theta_{2}=\mathbb{E}d_{\mathrm{H}}(X,W)$
where $d_{\mathrm{H}}(x,y)=\bone\{x\neq y\}$ denotes the Hamming
distance. If  $\mathbb{E}d_{\mathrm{H}}(X,W)>D_{1}$, then we choose
$\theta_{1}=D_{1}$; otherwise, we choose $\hat{X}=W$. We choose
$\hat{Y}$ in a similar way. This set of induced distributions $(P_{W|XY},P_{\hat{X}|WX},P_{\hat{Y}|WY})$
obviously satisfies the distortion constraints in \eqref{eq:lossyRD}.
Moreover, if $R_{1}\ge R(D_{1})$, then 
\begin{align*}
I(X;\hat{X}|W) & =H(X|W)-H(X|\hat{X})\\
 & \le I(X;\hat{X})=R(D_{1})\le R_{1}.
\end{align*}
If $R_{1}<R(D_{1})$, then $I(X;W)\ge R(D_{1})-R_{1}$ (since $P_{W|XY}$
attains $\Upsilon(\left[R(D_{1})-R_{1}\right]^{+},\left[R(D_{2})-R_{2}\right]^{+})$).
So, for this case, 
\begin{align*}
I(X;\hat{X}|W) & =I(X;\hat{X})-I(X;W)\\
 & =R(D_{1})-I(X;W)\le R_{1}.
\end{align*}
So, it always holds that $I(X;\hat{X}|W)\le R_{1}$ for any cases.
By symmetry, $I(Y;\hat{Y}|W)\le R_{2}$ also holds. So, $(P_{W|XY},P_{\hat{X}|WX},P_{\hat{Y}|WY})$
also satisfies the rate constraints in \eqref{eq:lossyRD}. This implies
that $(P_{W|XY},P_{\hat{X}|WX},P_{\hat{Y}|WY})$ is a feasible solution
to \eqref{eq:lossyRD}, and hence, 
\begin{align}
 & R_{0}(R_{1},R_{2},D_{1},D_{2})\nonumber \\
 & \le I(X,Y;W)\\
 & =\Upsilon(\left[R(D_{1})-R_{1}\right]^{+},\left[R(D_{2})-R_{2}\right]^{+}).\label{eq:-19}
\end{align}
This completes the proof. 
\end{IEEEproof}
\begin{rem}
\label{rem:layeredcoding}A more straightforward way to show the inequality
in \eqref{eq:-19} is to use a specific coding scheme in which the
sender first encodes the source into $W=(U,V)$ by using an optimal
point-to-point lossy compression code with distortions $(D_{1}',D_{2}')$
and common rate $R_{XY}(D_{1}',D_{2}')$ where $D_{1}'=R^{-1}\left(\left[R(D_{1})-R_{1}\right]^{+}\right),D_{2}'=R^{-1}\left(\left[R(D_{2})-R_{2}\right]^{+}\right)$,
and then further encodes $X$ and $Y$ with help of $U,V$ by using
successively refinement codes with private rates $h(D_{1}')-h(D_{1})$
and $h(D_{2}')-h(D_{2})$ respectively. Note that here
\begin{align*}
h(D_{1}')-h(D_{1}) & =R(D_{1})-R(D_{1}')\\
 & =R(D_{1})-\left[R(D_{1})-R_{1}\right]^{+}\\
 & =\min\{R_{1},R(D_{1})\}\\
 & \le R_{1},
\end{align*}
and similarly, $h(D_{1}')-h(D_{1})\le R_{2}$. This scheme is essentially
same as the lossless one given below Conjecture \ref{conj:-For-the}.
\end{rem}
We next derive analytical expressions for $\underline{\Upsilon}(\alpha,\beta)$
and $\overline{\Upsilon}(\alpha,\beta)$.  Define 
\begin{align*}
\mathcal{I}_{0}^{*} & :=\bigl\{(\alpha,\beta)\in[0,1]^{2}:\beta\ge\alpha-\alpha h(p/\alpha),\\
 & \qquad\qquad\alpha\ge\beta-\beta h(p/\beta)\bigr\}.
\end{align*}
Define 
\begin{align*}
\mathcal{D}_{3}' & :=\left\{ (\alpha,\beta)\in[0,1]^{2}:a*p<b,\;\beta\ge\left(1-h(p)\right)\alpha\right\} ,\\
\mathcal{D}_{4}' & :=\left\{ (\alpha,\beta)\in[0,1]^{2}:b*p<a,\;\alpha\ge\left(1-h(p)\right)\beta\right\} ,\\
\mathcal{D}_{3}'' & :=\bigl\{(\alpha,\beta)\in[0,1]^{2}:\alpha-\alpha h(p/\alpha)\le\beta\\
 & \qquad\qquad<\left(1-h(p)\right)\alpha\bigr\},\\
\mathcal{D}_{4}'' & :=\bigl\{(\alpha,\beta)\in[0,1]^{2}:\beta-\beta h(p/\beta)\le\alpha\\
 & \qquad\qquad<\left(1-h(p)\right)\beta\bigr\}.
\end{align*}
Then, $\mathcal{I}_{0}^{*}=\mathcal{D}_{1}\cup\mathcal{D}_{2}\cup\mathcal{D}_{3}'\cup\mathcal{D}_{3}''\cup\mathcal{D}_{4}'\cup\mathcal{D}_{4}''$.
For $(\alpha,\beta)\in\mathcal{I}_{0}^{*}$, define 
\begin{align}
 & \underline{\Upsilon}^{*}(\alpha,\beta)\nonumber \\
 & :=\begin{cases}
1-(1-p)h\left(\frac{a+b-p}{2(1-p)}\right)\\
\qquad-ph\left(\frac{a-b+p}{2p}\right), & (\alpha,\beta)\in\mathcal{D}_{1}\\
1+h(p)-h(a)-h(b), & (\alpha,\beta)\in\mathcal{D}_{2}\\
\alpha, & (\alpha,\beta)\in\mathcal{D}_{3}'\\
h(p)+\beta\\
\qquad-(1-\alpha)h\left(\frac{p-\alpha h^{-1}(1-\beta/\alpha)}{1-\alpha}\right), & (\alpha,\beta)\in\mathcal{D}_{3}''\\
\beta, & (\alpha,\beta)\in\mathcal{D}_{4}'\\
h(p)+\alpha\\
\qquad-(1-\beta)h\left(\frac{p-\beta h^{-1}(1-\alpha/\beta)}{1-\beta}\right), & (\alpha,\beta)\in\mathcal{D}_{4}''
\end{cases},\label{eq:mi-6}
\end{align}
where  $a=h^{-1}(1-\alpha),b=h^{-1}(1-\beta)$. For $(\alpha,\beta)\in\mathcal{I}_{0}^{*}$,
define 
\[
\overline{\Upsilon}^{*}(\alpha,\beta):=h(p)+\alpha\land\beta.
\]

We now provide analytical expressions for $\underline{\Upsilon}(\alpha,\beta)$
and $\overline{\Upsilon}(\alpha,\beta)$ in the following theorem.
Since $\Upsilon$ is determined by $\underline{\Upsilon}$, this theorem
can be seen as an improved version of Theorem \ref{thm:mi}.   
\begin{thm}[Mutual Information Region for DSBS]
 \label{thm:mi2} For the source $\ensuremath{\DSBS(p)}$ with $p\in(0,1/2)$,
 the following hold. 
\begin{enumerate}
\item The projection region satisfies 
\[
\mathcal{I}_{0}=\mathcal{I}_{0}^{*}.
\]
\item For $(\alpha,\beta)\in\mathcal{I}_{0}$, the lower and upper envelopes
of the mutual information region satisfy 
\begin{align}
\underline{\Upsilon}(\alpha,\beta) & =\underline{\Upsilon}^{*}(\alpha,\beta),\label{eq:mi-2}\\
\overline{\Upsilon}(\alpha,\beta) & =\overline{\Upsilon}^{*}(\alpha,\beta).
\end{align}
\end{enumerate}
\end{thm}

The proof is provided in Section \ref{sec:Proof-of-Theorem-mi-1}
which follows steps same as those for Theorem \ref{thm:mi}. Note
that $\Upsilon^{*}$ and $\underline{\Upsilon}^{*}$ differ on the
regions $\mathcal{D}_{3}''$ and $\mathcal{D}_{4}''.$ The functions
$\underline{\Upsilon}^{*}$ and $\overline{\Upsilon}^{*}$ are plotted
in Fig. \ref{fig:upsilon}.

\subsection{Mutual Information Region for Gaussian Source }

We next consider Gaussian sources. In this subsection and in the
corresponding proofs of results stated in this subsection, we always
use the logarithm with base $e$, which is denoted by $\ln$. 

Let $P_{XY}=\mathcal{N}(\boldsymbol{0},\boldsymbol{\Sigma})$ with
\[
\boldsymbol{\Sigma}=\begin{bmatrix}1 & \rho\\
\rho & 1
\end{bmatrix}
\]
and $\rho\in(0,1)$. We next give the analytical expression for the
mutual information region for a Gaussian source. For $\alpha\ge0$,
denote $\theta_{\alpha}\in[0,\pi/2]$ such that
\begin{align*}
\sin\theta_{\alpha} & =e^{-\alpha}.
\end{align*}
So, we also have $\sin\theta_{\beta}=e^{-\beta}.$ 

Denote 
\begin{align}
\rho_{\alpha,\beta} & :=\frac{\rho-\sqrt{\left(1-e^{-2\alpha}\right)\left(1-e^{-2\beta}\right)}}{e^{-\alpha-\beta}}\nonumber \\
 & =\frac{\rho-\cos\theta_{\alpha}\cos\theta_{\beta}}{\sin\theta_{\alpha}\sin\theta_{\beta}}.\label{eq:rhohat}
\end{align}
Define several disjoint sets 
\begin{align*}
\mathcal{D}_{\mathrm{G},1} & :=\Bigl\{(\alpha,\beta)\in[0,\infty)^{2}:\cos\theta_{\alpha}\cos\theta_{\beta}\le\rho\\
 & \qquad\qquad\le\min\left\{ \frac{\cos\theta_{\beta}}{\cos\theta_{\alpha}},\frac{\cos\theta_{\alpha}}{\cos\theta_{\beta}}\right\} \Bigr\},\\
\mathcal{D}_{\mathrm{G},2} & :=\Bigl\{(\alpha,\beta)\in[0,\infty)^{2}:\\
 & \qquad\rho\le\min\left\{ \cos\theta_{\alpha}\cos\theta_{\beta},\frac{\cos\theta_{\beta}}{\cos\theta_{\alpha}},\frac{\cos\theta_{\alpha}}{\cos\theta_{\beta}}\right\} \Bigr\},\\
\mathcal{D}_{\mathrm{G},3} & :=\left\{ (\alpha,\beta)\in[0,\infty)^{2}:\rho>\frac{\cos\theta_{\beta}}{\cos\theta_{\alpha}}\right\} ,\\
\mathcal{D}_{\mathrm{G},4} & :=\left\{ (\alpha,\beta)\in[0,\infty)^{2}:\rho>\frac{\cos\theta_{\alpha}}{\cos\theta_{\beta}}\right\} .
\end{align*}
Define a function for $\alpha,\beta\ge0$, 
\begin{equation}
\Upsilon_{\mathrm{G}}^{*}(\alpha,\beta):=\begin{cases}
\alpha+\beta-\frac{1}{2}\ln\frac{1-\rho_{\alpha,\beta}^{2}}{1-\rho^{2}} & (\alpha,\beta)\in\mathcal{D}_{\mathrm{G},1}\\
\alpha+\beta-\frac{1}{2}\ln\frac{1}{1-\rho^{2}} & (\alpha,\beta)\in\mathcal{D}_{\mathrm{G},2}\\
\alpha & (\alpha,\beta)\in\mathcal{D}_{\mathrm{G},3}\\
\beta & (\alpha,\beta)\in\mathcal{D}_{\mathrm{G},4}
\end{cases}.\label{eq:G-1}
\end{equation}

\begin{thm}[Mutual Information Region for Gaussian Source]
 \label{thm:miGaussian} For the bivariate Gaussian source $\mathcal{N}(\boldsymbol{0},\boldsymbol{\Sigma})$,
 the following hold. 
\begin{enumerate}
\item The projection region satisfies $\mathcal{I}_{0}=[0,\infty)^{2}.$
\item For $(\alpha,\beta)\in\mathcal{I}_{0}$, the lower and upper envelopes
of the mutual information region satisfy 
\begin{align}
\underline{\Upsilon}(\alpha,\beta) & =\Upsilon_{\mathrm{G}}^{*}(\alpha,\beta),\label{eq:mi-7}\\
\overline{\Upsilon}(\alpha,\beta) & =+\infty.\label{eq:mi-8}
\end{align}
Moreover, $\Upsilon_{\mathrm{G}}^{*}$ is increasing in one parameter
given the other one, and hence, the lower increasing envelope satisfies
$\Upsilon(\alpha,\beta)=\Upsilon_{\mathrm{G}}^{*}(\alpha,\beta)$.
\end{enumerate}
\end{thm}
The proof of Theorem \ref{thm:miGaussian} is provided in Section
\ref{sec:Proof-of-Theorem-Gaussian}, which is similar to those of
Theorems \ref{thm:mi} and \ref{thm:mi2}. More specifically, it is
based on an auxiliary measure technique and the analytical expression
of the optimal-transport divergence region for the Gaussian source.
The analytical expression of the optimal-transport divergence region
for the Gaussian source is given in Lemma \ref{lem:convexity-Gaussian}
in Section \ref{sec:Preliminaries-on-Optimal-Transpo}. Another possible
way to prove Theorem \ref{thm:miGaussian} is based on the fact \cite{nair2014extremal,liu2018information}
that it suffices to evaluate the mutual information region for the
Gaussian source by using a random variable $W$ which is jointly Gaussian
with $X,Y$. By this fact, evaluating the mutual information region
over arbitrary auxiliary random variable $W$ reduces to evaluating
it over the covariance matrix of $(W,X,Y)$ and the mean of $W$.
Note that the resultant optimization is still nonconvex, and hence,
solving it requires some additional techniques.

The function $\Upsilon_{\mathrm{G}}^{*}$ is plotted in Fig. \ref{fig:upsilon}.
By the following lemma, it holds that $\cos\left(\theta_{\beta}-\theta_{\alpha}\right)\ge\min\left\{ \frac{\cos\theta_{\beta}}{\cos\theta_{\alpha}},\frac{\cos\theta_{\alpha}}{\cos\theta_{\beta}}\right\} $,
which implies $0\le\rho_{\alpha,\beta}<1$ for $(\alpha,\beta)\in\mathcal{D}_{\mathrm{G},1}$. 
\begin{lem}
For $0\le\theta_{1}\le\theta_{2}<\pi/2$, it holds that $\cos\theta_{2}\le\cos\left(\theta_{2}-\theta_{1}\right)\cos\theta_{1}$.
\end{lem}
\begin{IEEEproof}
This lemma is obviously since $\cos\theta_{2}=\cos\left(\theta_{2}-\theta_{1}\right)\cos\theta_{1}-\sin\left(\theta_{2}-\theta_{1}\right)\sin\theta_{1}\le\cos\left(\theta_{2}-\theta_{1}\right)\cos\theta_{1}$. 
\end{IEEEproof}

\begin{figure}
\centering \includegraphics[width=0.45\textwidth]{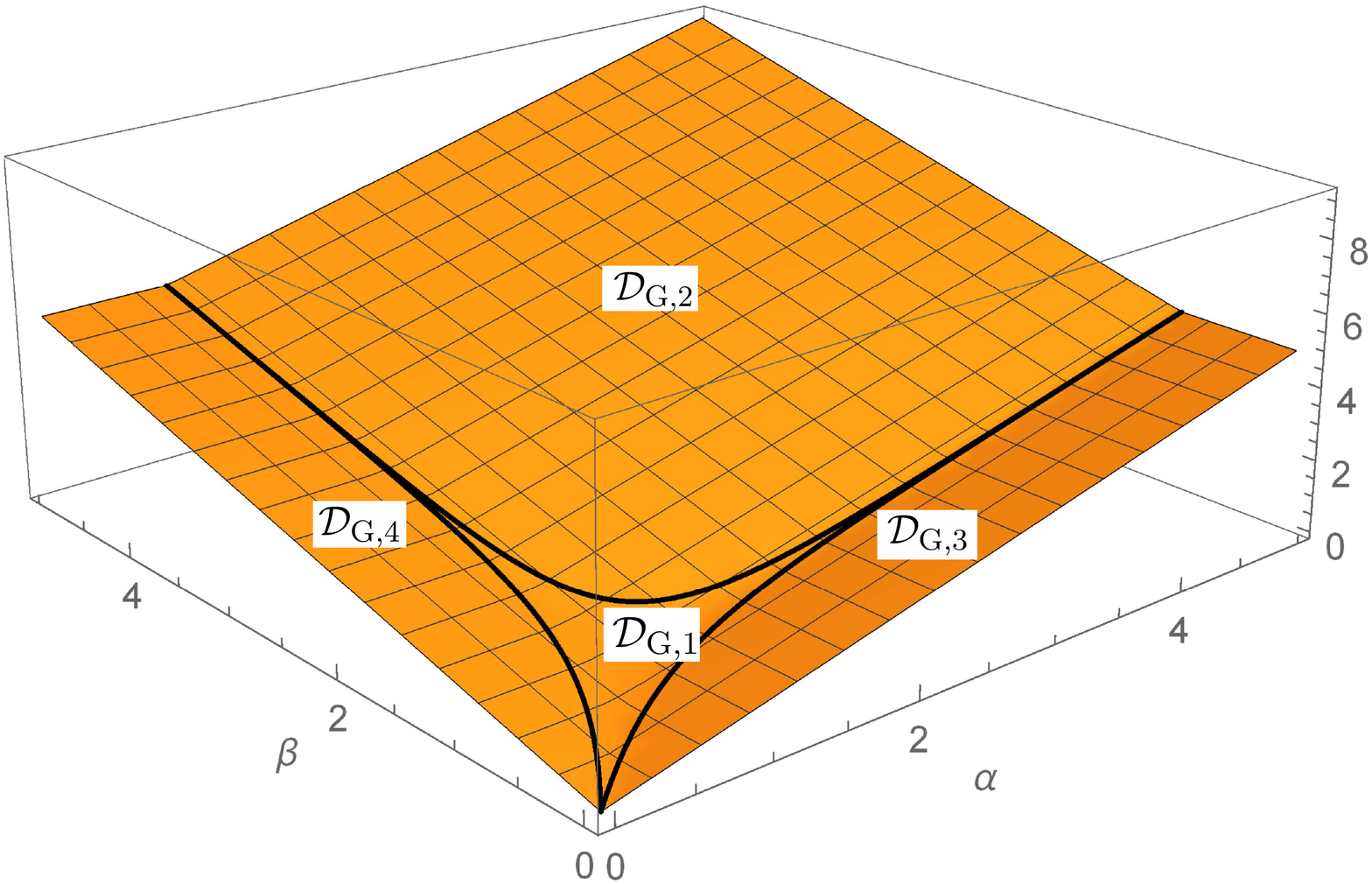}

\caption{\label{fig:upsilon-1}Illustration of $\Upsilon_{\mathrm{G}}^{*}$
for $\rho=0.9$. }
\end{figure}

Using Theorem \ref{thm:miGaussian}, we obtain the analytical expression
for the lossy Gray--Wyner rate region of a Gaussian source. The proof
is similar to that of Corollary \ref{cor:lossyGW}, and hence, omitted
here. 
\begin{cor}[Lossy Gray--Wyner Rate Region for Gaussian Source]
\label{cor:Gaussian} For the bivariate Gaussian source $\mathcal{N}(\boldsymbol{0},\boldsymbol{\Sigma})$,
under the quadratic distortion measure, it holds that for $R_{1},R_{2},D_{1},D_{2}\ge0$,
\begin{align*}
 & R_{0}(R_{1},R_{2},D_{1},D_{2})\\
 & =\Upsilon_{\mathrm{G}}^{*}(\left[R_{\mathrm{G}}(D_{1})-R_{1}\right]^{+},\left[R_{\mathrm{G}}(D_{2})-R_{2}\right]^{+}),
\end{align*}
where $\Upsilon_{\mathrm{G}}^{*}$ is defined in \eqref{eq:G-1},
$\left[x\right]^{+}:=\max\{x,0\}$, and $R_{\mathrm{G}}(D)=\frac{1}{2}\ln\frac{1}{D}$
is the rate-distribution function of the standard Gaussian source
$\mathcal{N}(0,1)$. 
\end{cor}

A partial result of Corollary \ref{cor:Gaussian} was given in \cite{sula2021gray},
where the analytical expression for the function $R_{0}(R_{\mathrm{s}},D_{1},D_{2}):=\min_{R_{1}+R_{2}=R_{\mathrm{s}}}R_{0}(R_{1},R_{2},D_{1},D_{2})$
was derived. Furthermore, Corollary \ref{cor:Gaussian} in a different
form was presented in \cite{chen2021computing} by using a different
method. 

Note that $R_{\mathrm{G}}(D_{1},D_{2})=\Upsilon_{\mathrm{G}}^{*}(R_{\mathrm{G}}^{-1}(D_{1}),R_{\mathrm{G}}^{-1}(D_{2}))$
 where $R_{\mathrm{G}}(\cdot,\cdot)$ is the rate-distortion function
for the bivariate Gaussian source $\mathcal{N}(\boldsymbol{0},\boldsymbol{\Sigma})$
under the quadratic distortion measure \cite{Berger,xiao2005compression}.
Corollary \ref{cor:Gaussian} implicitly states that a layered coding
scheme similar to the one given in Remark \ref{rem:layeredcoding}
is optimal for the lossy Gray--Wyner system for the bivariate Gaussian
source. 

\subsection{Implications of Our Results}

The Gray--Wyner rate region has many applications. It has not only
been used to characterize the rate region of the Gray--Wyner coding
system, but also used to characterize many other problems, including
the measure of common information \cite{WynerCI}, the exponent of
the maximal density of the type graph \cite{yu2021Graphs}, the optimal
exponent in the Brascamp--Lieb (BL) inequalities for uniform distributions
over type classes \cite{yu2021Graphs,liu2019smoothing}, the hypercontractivity
region \cite{ahlswede1976spreading,anantharam2014hypercontractivity,nair2014equivalent},
Mrs. Gerber's lemma and information bottleneck \cite{wyner1973theorem,witsenhausen1974entropy,tishby2000information},
communication rate for channel synthesis \cite{cuff13}, etc. See
more details in \cite{li2017extended}. So, our characterizations
of the Gray--Wyner rate regions for the DSBS and the Gaussian source
imply the corresponding characterizations of these results for the
same sources, although some of them are already known.

Furthermore, in theoretical computer science, the DSBS is usually
described as a coin toss model. Such a source has now attracted a
lot of interest in theoretical computer science. For example, the
joint probability of $A\times A$ under the DSBS corresponds to the
generating function of the Fourier weights of the Boolean function
$\bone_{A}$, and hence, the DSBS (and also its hypercontractivity
inequalities) plays a key role  in analysis of Boolean functions;
see, e.g., \cite{ODonnell14analysisof} for more details.

\section{\label{sec:Preliminaries-on-Optimal-Transpo}Preliminaries on Optimal-Transport
Divergences}

Before proving the main results, we first introduce some preliminary
lemmas that will be used in our proofs. 

The set of all couplings with marginals $Q_{X}$ and $Q_{Y}$ is
denoted as 
\begin{align*}
 & \calC(P_{X},P_{Y})\\
 & :=\big\{ Q_{XY}\in\calP(\calX\times\calY):Q_{X}=P_{X},Q_{Y}=P_{Y}\big\}.
\end{align*}

\begin{defn}
The optimal transport divergence (or minimum relative entropy) between
$Q_{X}$ and $Q_{Y}$ with respect to a probability measure $P_{XY}$
is defined as 
\begin{equation}
\D(Q_{X},Q_{Y}\|P_{XY}):=\inf_{Q_{XY}\in\calC(Q_{X},Q_{Y})}D(Q_{XY}\|P_{XY}).\label{eq:mathbbD}
\end{equation}
\end{defn}
Define the optimal-transport-divergence (or minimum-relative-entropy)
region of $P_{XY}$ as 
\begin{align*}
\mathcal{D}\left(P_{XY}\right) & :=\bigcup_{Q_{X}\ll P_{X},Q_{Y}\ll P_{Y}}\bigl\{(D(Q_{X}\|P_{X}),D(Q_{Y}\|P_{Y}),\\
 & \qquad\qquad\D(Q_{X},Q_{Y}\|P_{XY}))\bigr\}.
\end{align*}
Define the lower and upper envelopes of the optimal divergence region
$\mathcal{D}\left(P_{XY}\right)$ as for $\alpha,\beta\ge0$, 
\begin{align}
\underline{\varphi}(\alpha,\beta) & :=\inf_{\substack{Q_{XY}:D(Q_{X}\|P_{X})=\alpha,\\
D(Q_{Y}\|P_{Y})=\beta
}
}D(Q_{XY}\|P_{XY})\label{eq:-21}\\
 & =\inf_{\substack{Q_{X},Q_{Y}:D(Q_{X}\|P_{X})=\alpha,\\
D(Q_{Y}\|P_{Y})=\beta
}
}\D(Q_{X},Q_{Y}\|P_{XY})\label{eq:-20}
\end{align}
and 
\begin{align}
\overline{\varphi}(\alpha,\beta) & :=\sup_{\substack{Q_{X},Q_{Y}:D(Q_{X}\|P_{X})=\alpha,\\
D(Q_{Y}\|P_{Y})=\beta
}
}\D(Q_{X},Q_{Y}\|P_{XY})\label{eq:-22}
\end{align}
Define the lower and upper increasing envelopes of $\mathcal{D}\left(P_{XY}\right)$
respectively as 
\begin{align}
\underline{\psi}(\alpha,\beta) & :=\inf_{s\ge\alpha,t\ge\beta}\underline{\varphi}(s,t)\label{eq:lce-1}\\
\overline{\psi}(\alpha,\beta) & :=\sup_{s\leq\alpha,t\leq\beta}\overline{\varphi}(s,t).\label{eq:uce-1}
\end{align}
We also define for $q<0$, 
\begin{align*}
\varphi_{q}(\alpha) & :=\sup_{Q_{X}:D(Q_{X}\|P_{X})=\alpha}\inf_{Q_{Y}}\D(Q_{X},Q_{Y}\|P_{XY})\\
 & \qquad\qquad-\frac{D(Q_{Y}\|P_{Y})}{q}.
\end{align*}
For the DSBS, $\varphi_{q}$ can be rewritten as 
\begin{align}
\varphi_{q}(\alpha) & =\min_{0\le\beta\le1}\underline{\varphi}(\alpha,\beta)-\frac{\beta}{q}.\label{eq:phi_q-2}
\end{align}

\begin{figure*}
\centering %
\begin{tabular}{cc}
\includegraphics[width=0.45\textwidth]{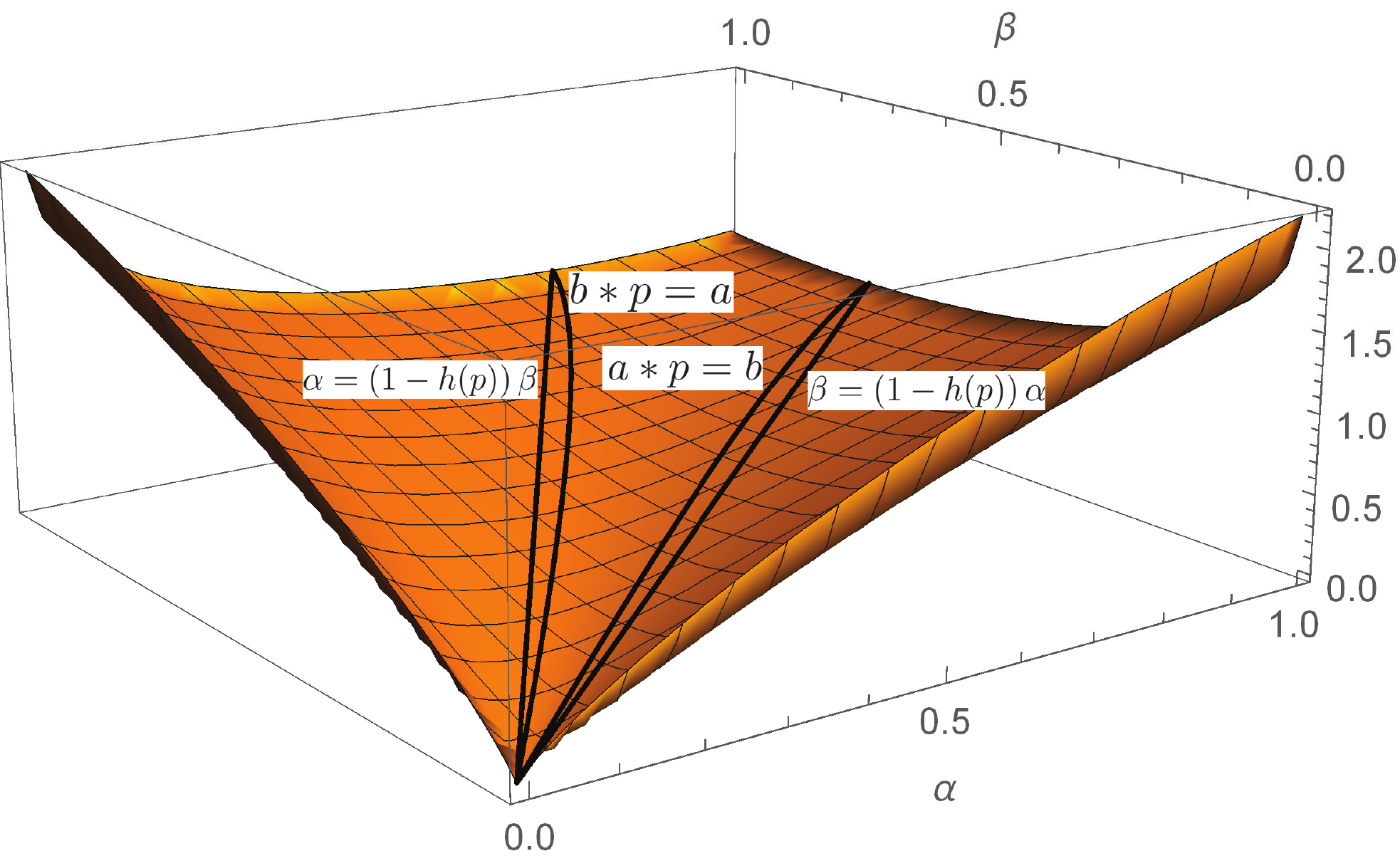} & \includegraphics[width=0.45\textwidth]{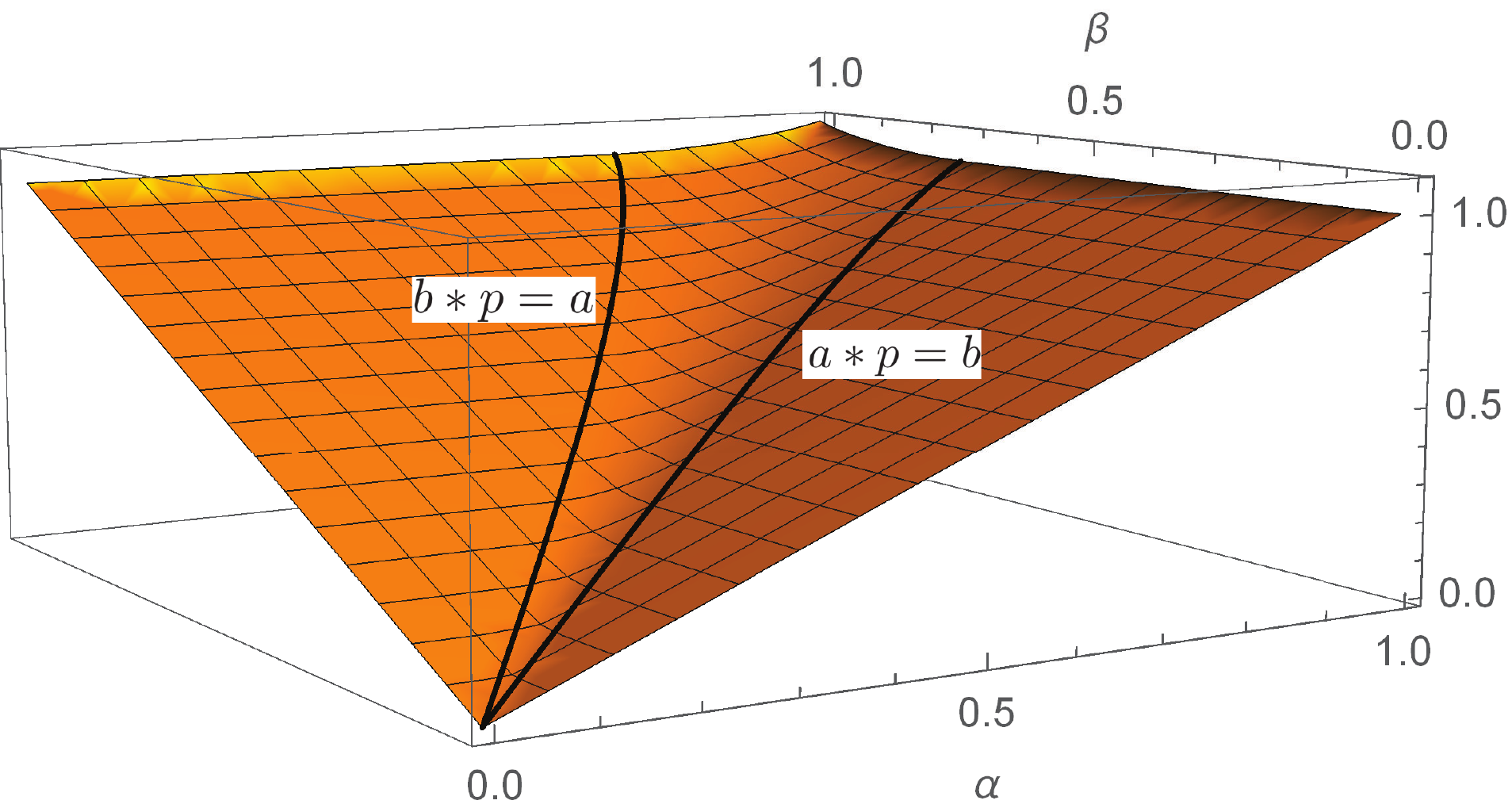}\tabularnewline
{\footnotesize{}(a) $\underline{\varphi}$} & {\footnotesize{}(b) $\underline{\psi}$}\tabularnewline
\end{tabular}

\vspace{.2cm}

\centering%
\begin{tabular}{cc}
\includegraphics[width=0.45\textwidth]{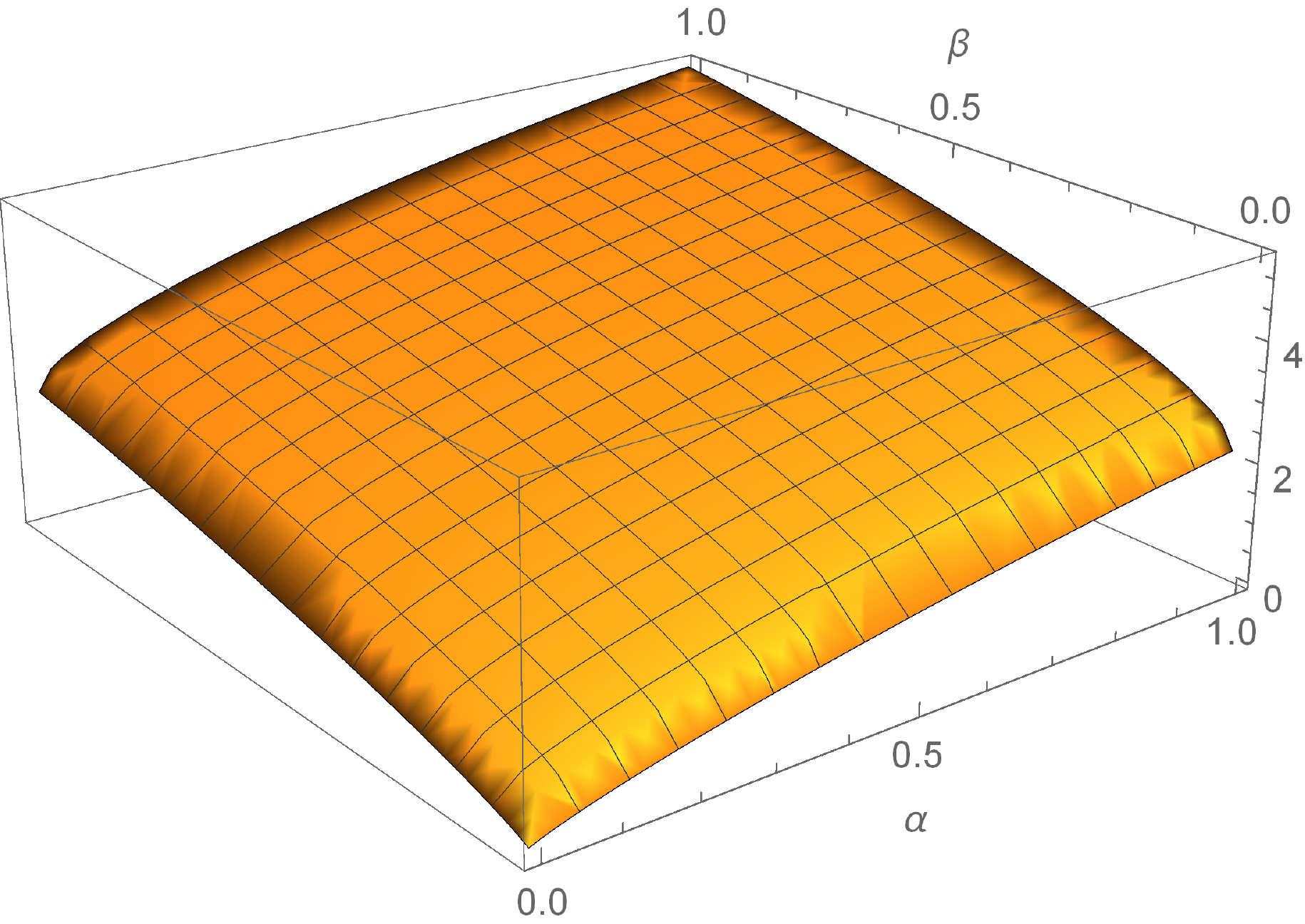} & \includegraphics[width=0.45\textwidth]{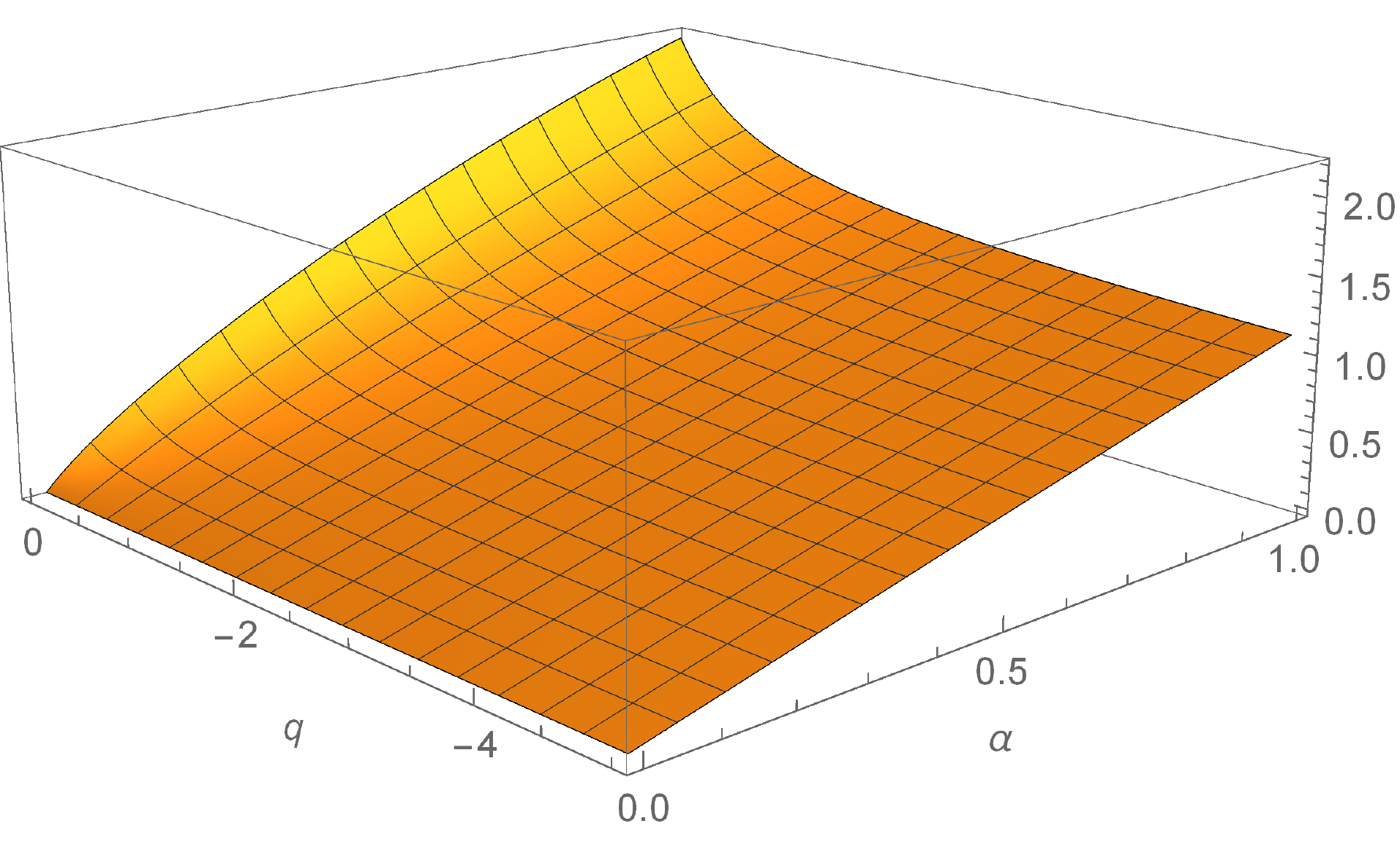}\tabularnewline
{\footnotesize{}(c) $\overline{\psi}=\overline{\varphi}$} & {\footnotesize{}(d) $\varphi_{q}$}\tabularnewline
\end{tabular}

\caption{\label{fig:phi}Illustration of $\underline{\varphi},\underline{\psi}$,
$\overline{\psi}=\overline{\varphi}$, and $\varphi_{q}$ for the
$\ensuremath{\DSBS(p)}$ with $p=0.05$ (equivalently, the correlation
coefficient $\rho=0.9$). Lemma \ref{lem:convexity} implies that
$\underline{\psi}$ is convex, $\overline{\psi}=\overline{\varphi}$
is concave, and $\varphi_{q}$ is convex for $q<0$.}
\end{figure*}

The following lemma is obvious. Recall that, as mentioned in the
notation part (at the end of the introduction section), $\conv f$
and $\conc f$ respectively denote the lower convex envelope and the
upper concave envelope of $f$. 
\begin{lem}
\label{lem:timesharing} It holds that 
\begin{align}
 & \conv\underline{\psi}(\alpha,\beta)\nonumber \\
 & =\inf_{s\ge\alpha,t\ge\beta}\conv\underline{\varphi}(s,t)\\
 & =\inf_{\substack{Q_{UXY}:\\
D(Q_{X|U}\|P_{X}|Q_{U})\ge\alpha,\\
D(Q_{Y|U}\|P_{Y}|Q_{U})\ge\beta
}
}D(Q_{XY|U}\|P_{XY}|Q_{U}),\label{eq:-25}
\end{align}
and 
\begin{align}
 & \conc\overline{\psi}(\alpha,\beta)\nonumber \\
 & =\sup_{s\leq\alpha,t\leq\beta}\conc\overline{\psi}(s,t)\\
 & =\sup_{\substack{Q_{U},Q_{X|U},Q_{Y|U}:\\
D(Q_{X|U}\|P_{X}|Q_{U})\le\alpha,\\
D(Q_{Y|U}\|P_{Y}|Q_{U})\le\beta
}
}\D(Q_{X|U},Q_{Y|U}\|P_{XY}|Q_{U}),\label{eq:-25-1}
\end{align}
where 
\begin{align*}
 & \D(Q_{X|U},Q_{Y|U}\|P_{XY}|Q_{U})\\
 & :=\inf_{Q_{XY|U}\in\calC(Q_{X|U},Q_{Y|U})}D(Q_{XY|U}\|P_{XY}|Q_{U})
\end{align*}
with $\calC(Q_{X|U},Q_{Y|U})$ denoting the set of conditional distributions
$Q_{XY|U}$ whose marginals are $Q_{X|U},Q_{Y|U}$. Furthermore, the
alphabet sizes of $U$ in the last infimization in \eqref{eq:-25}
and the last supremization in \eqref{eq:-25-1} can be restricted
to be no larger than $3$. 
\end{lem}
\begin{IEEEproof}
The bound on the alphabet sizes of $U$ follows by the support lemma
\cite{elgamal}. Based on this, it is easily seen that the last infimization
in \eqref{eq:-25} is equal to 
\begin{align}
 & \inf_{\substack{(q_{i},Q_{XY}^{(i)})_{i\in[3]}:\\
\sum_{i}q_{i}=1,\;q_{i}\ge0,\forall i\in[3]\\
\sum_{i}q_{i}D(Q_{X}^{(i)}\|P_{X})\ge\alpha,\\
\sum_{i}q_{i}D(Q_{Y}^{(i)}\|P_{Y})\ge\beta
}
}\sum_{i}q_{i}D(Q_{XY}^{(i)}\|P_{XY})\nonumber \\
 & =\inf_{\substack{(q_{i},s_{i},t_{i})_{i\in[3]}:\\
\sum_{i}q_{i}=1,\;q_{i}\ge0,\forall i\in[3]\\
\sum_{i}q_{i}s_{i}\ge\alpha,\sum_{i}q_{i}t_{i}\ge\beta
}
}\inf_{\substack{(Q_{XY}^{(i)})_{i\in[3]}:\\
D(Q_{X}^{(i)}\|P_{X})=s_{i},\\
D(Q_{Y}^{(i)}\|P_{Y})=t_{i}
}
}\sum_{i}q_{i}D(Q_{XY}^{(i)}\|P_{XY})\label{eq:-27}\\
 & =\inf_{\substack{(q_{i},s_{i},t_{i})_{i\in[3]}:\\
\sum_{i}q_{i}=1,\;q_{i}\ge0,\forall i\in[3]\\
\sum_{i}q_{i}s_{i}\ge\alpha,\sum_{i}q_{i}t_{i}\ge\beta
}
}\sum_{i}q_{i}\inf_{\substack{Q_{XY}^{(i)}:\\
D(Q_{X}^{(i)}\|P_{X})=s_{i},\\
D(Q_{Y}^{(i)}\|P_{Y})=t_{i}
}
}D(Q_{XY}^{(i)}\|P_{XY})\label{eq:-28}\\
 & =\inf_{\substack{(q_{i},s_{i},t_{i})_{i\in[3]}:\\
\sum_{i}q_{i}=1,\;q_{i}\ge0,\forall i\in[3]\\
\sum_{i}q_{i}s_{i}\ge\alpha,\sum_{i}q_{i}t_{i}\ge\beta
}
}\sum_{i}q_{i}\underline{\varphi}(s_{i},t_{i}),\nonumber 
\end{align}
where $(q_{i})_{i\in[3]}$ denotes the probability values of $U$,
$Q_{XY}^{(i)}$ denotes $Q_{XY|U=i}$, and \eqref{eq:-28} follows
since the inner infimization in \eqref{eq:-27} can be taken pointwise
for each $i$. By definition, it is easily verified that both $\conv\underline{\psi}(\alpha,\beta)$
and $\inf_{s\ge\alpha,t\ge\beta}\conv\underline{\varphi}(s,t)$ are
also equal to the last formula above. So, equalities in \eqref{eq:-25}
hold. Similarly, one can prove that equalities in \eqref{eq:-25-1}
hold as well. 

\end{IEEEproof}
The analytic expressions for various envelopes of the optimal divergence
region for the DSBS are given in the following lemma.  
\begin{lem}
\cite{yu2021convexity_article} \label{lem:convexity} For the source
$\ensuremath{\DSBS(p)}$ with $p\in(0,1/2)$,  the following hold. 
\begin{enumerate}
\item It holds that for $\alpha,\beta\in[0,1]^{2}$, the optimal distribution
$Q_{XY}$ attaining $\underline{\varphi}(\alpha,\beta)$ (in \eqref{eq:-21})
is
\[
Q_{XY}=\begin{bmatrix}1+q-a-b & b-q\\
a-q & q
\end{bmatrix},
\]
 where $a=h^{-1}(1-\alpha),b=h^{-1}(1-\beta)$, and 
\begin{align*}
q & =q_{a,b}(p):=\frac{1}{2\left(\kappa-1\right)}\times\Bigl(\left(\kappa-1\right)\left(a+b\right)+1\\
 & \qquad-\sqrt{\left(\left(\kappa-1\right)\left(a+b\right)+1\right)^{2}-4\kappa\left(\kappa-1\right)ab}\Bigr)
\end{align*}
with $\kappa=\left(\frac{1-p}{p}\right)^{2}$. Similarly, the optimal
distribution $Q_{XY}$ attaining $\overline{\varphi}(\alpha,\beta)$
(in \eqref{eq:-22}) is still $Q_{XY}$ but with $b$ replaced by
$b=1-h^{-1}(1-\beta)$ (or alternatively, with $a$ replaced by $a=1-h^{-1}(1-\alpha)$).
\item Given $\alpha\in[0,1]$, $\beta\mapsto\underline{\varphi}(\alpha,\beta)$
is strictly decreasing for $\beta$ such that $a*p\le b$ and strictly
increasing for $\beta$ such that $a*p\ge b$, and moreover, its minimum
is $\alpha$ which is attained by the $\beta$ such that $a*p=b$.
Symmetrically, given $\beta\in[0,1]$, $\alpha\mapsto\underline{\varphi}(\alpha,\beta)$
is strictly decreasing for $\alpha$ such that $b*p\le a$ and strictly
increasing for $\alpha$ such that $b*p\ge a$, and moreover, its
minimum is $\beta$ which is attained by the $\alpha$ such that $b*p=a$.
\item It holds that for $\alpha,\beta\in[0,1]^{2}$, 
\begin{equation}
\underline{\psi}(\alpha,\beta)=\begin{cases}
\underline{\varphi}(\alpha,\beta) & a*p\ge b,\;b*p\ge a\\
\alpha & a*p<b\\
\beta & b*p<a
\end{cases}.\label{eq:exprphi}
\end{equation}
Moreover, $\underline{\psi}$ is convex on $[0,1]^{2}$ and strictly
convex on $\left\{ (\alpha,\beta):a*p\ge b,\;b*p\ge a\right\} $,
where $a=h^{-1}(1-\alpha),b=h^{-1}(1-\beta)$. 
\item It holds that for $\alpha,\beta\in[0,1]^{2}$, 
\begin{align}
 & \conv\underline{\varphi}(\alpha,\beta)\nonumber \\
 & =\begin{cases}
\underline{\varphi}(\alpha,\beta), & (\alpha,\beta)\in\hat{\mathcal{D}}_{1}\\
\alpha, & (\alpha,\beta)\in\hat{\mathcal{D}}_{2}\\
\alpha+\alpha D\bigl((1-h^{-1}(1-\beta/\alpha),\\
\quad h^{-1}(1-\beta/\alpha))\|(1-p,p)\bigr), & (\alpha,\beta)\in\hat{\mathcal{D}}_{4}\\
\beta, & (\alpha,\beta)\in\hat{\mathcal{D}}_{3}\\
\beta+\beta D\bigl((1-h^{-1}(1-\alpha/\beta),\\
\quad h^{-1}(1-\alpha/\beta))\|(1-p,p)\bigr), & (\alpha,\beta)\in\hat{\mathcal{D}}_{5}
\end{cases}\label{eq:exprphi-3}
\end{align}
where 
\begin{align*}
\hat{\mathcal{D}}_{1} & :=\mathcal{D}_{1}\cup\mathcal{D}_{2}=\bigl\{(\alpha,\beta)\in[0,1]^{2}:\\
 & \qquad\qquad a*p\ge b,\;b*p\ge a\bigr\},\\
\hat{\mathcal{D}}_{2} & :=\mathcal{D}_{3}'=\bigl\{(\alpha,\beta)\in[0,1]^{2}:\\
 & \qquad\qquad a*p<b,\;\beta\ge\left(1-h(p)\right)\alpha\bigr\},\\
\hat{\mathcal{D}}_{3} & :=\mathcal{D}_{4}'=\bigl\{(\alpha,\beta)\in[0,1]^{2}:\\
 & \qquad\qquad b*p<a,\;\alpha\ge\left(1-h(p)\right)\beta\bigr\},\\
\hat{\mathcal{D}}_{4} & :=\left\{ (\alpha,\beta)\in[0,1]^{2}:\beta<\left(1-h(p)\right)\alpha\right\} ,\\
\hat{\mathcal{D}}_{5} & :=\left\{ (\alpha,\beta)\in[0,1]^{2}:\alpha<\left(1-h(p)\right)\beta\right\} .
\end{align*}
Moreover, $\conv\underline{\varphi}(\alpha,\beta)$ for the second
clause above is attained by a convex combination of $(0,0)$ (with
probability $1-\theta$)  and $(1-h(a'),1-h(a'*p))$ (with probability
$\theta$), where $a'\in[0,1/2]$ is the unique solution to the equation
$\frac{1-h(a'*p)}{1-h(a')}=\frac{\beta}{\alpha}$, and $\theta=\frac{\alpha}{1-h(a')}$.
Similarly, $\conv\underline{\varphi}(\alpha,\beta)$ for the third
clause above is attained by a convex combination of $(0,0)$ (with
probability $1-\alpha$)  and $(1,\beta/\alpha)$ (with probability
$\alpha$). 
\item It holds that $\overline{\varphi}$ is increasing in one argument
given the other one. Moreover, $\overline{\varphi}$ is strictly
concave on $[0,1]^{2}$. 
\item For $q<0$, $\varphi_{q}$ is increasing and strictly concave on $[0,1]$.
\end{enumerate}
\end{lem}
\begin{rem}
\label{rem:optimaldistribution} In other words, the optimal distribution
$Q_{WXY}$ (with $W$ denoting the time-sharing random variable in
the convex combination operation) attaining $\conv\underline{\varphi}(\alpha,\beta)$
for the second clause in \eqref{eq:exprphi-3} is given by $Q_{W}=\mathrm{Bern}(\theta)$,
$Q_{XY|W=0}=P_{XY}$, and $Q_{XY|W=1}=\mathrm{Bern}(a')P_{Y|X}$;
the optimal distribution $Q_{WXY}$ attaining $\conv\underline{\varphi}(\alpha,\beta)$
for the third clause in \eqref{eq:exprphi-3} is given by $Q_{W}=\mathrm{Bern}(\alpha)$,
$Q_{XY|W=0}=P_{XY},$ and 
\begin{align*}
Q_{XY|W=1} & =\begin{bmatrix}1-h^{-1}(1-\beta/\alpha) & h^{-1}(1-\beta/\alpha)\\
0 & 0
\end{bmatrix}.
\end{align*}
\end{rem}
The functions appearing in Lemma \ref{lem:convexity} are plotted
in Fig. \ref{fig:phi}. All statements in Lemma \ref{lem:convexity}
were proven in \cite{yu2021convexity_article} except for Statements
2 and 4. The proofs of Statements 2 and 4 are given in Appendix \ref{sec:Proof-of-Lemma-1}. 

\begin{figure*}
\centering %
\begin{tabular}{cc}
\includegraphics[width=0.45\textwidth]{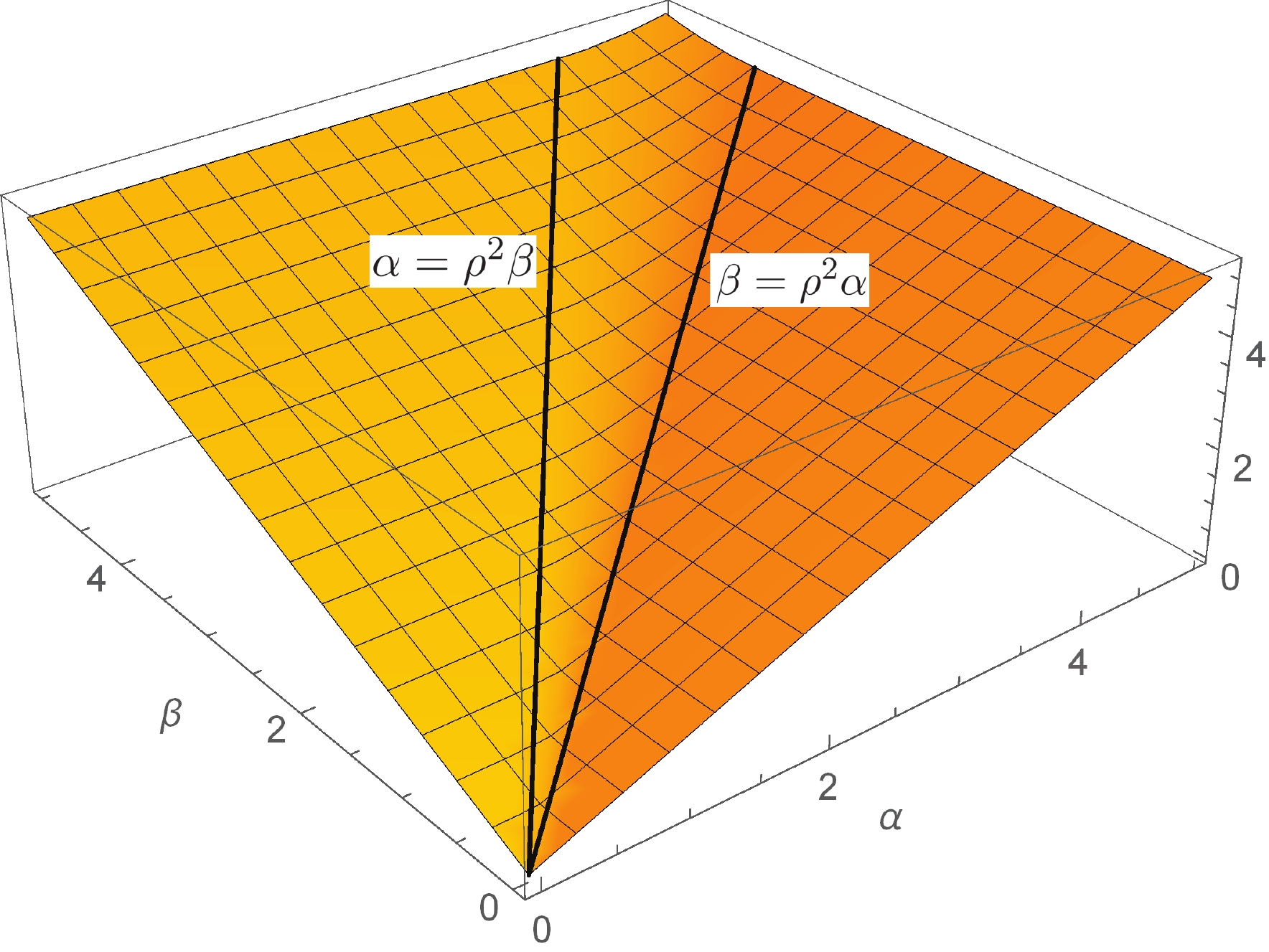} & \includegraphics[width=0.45\textwidth]{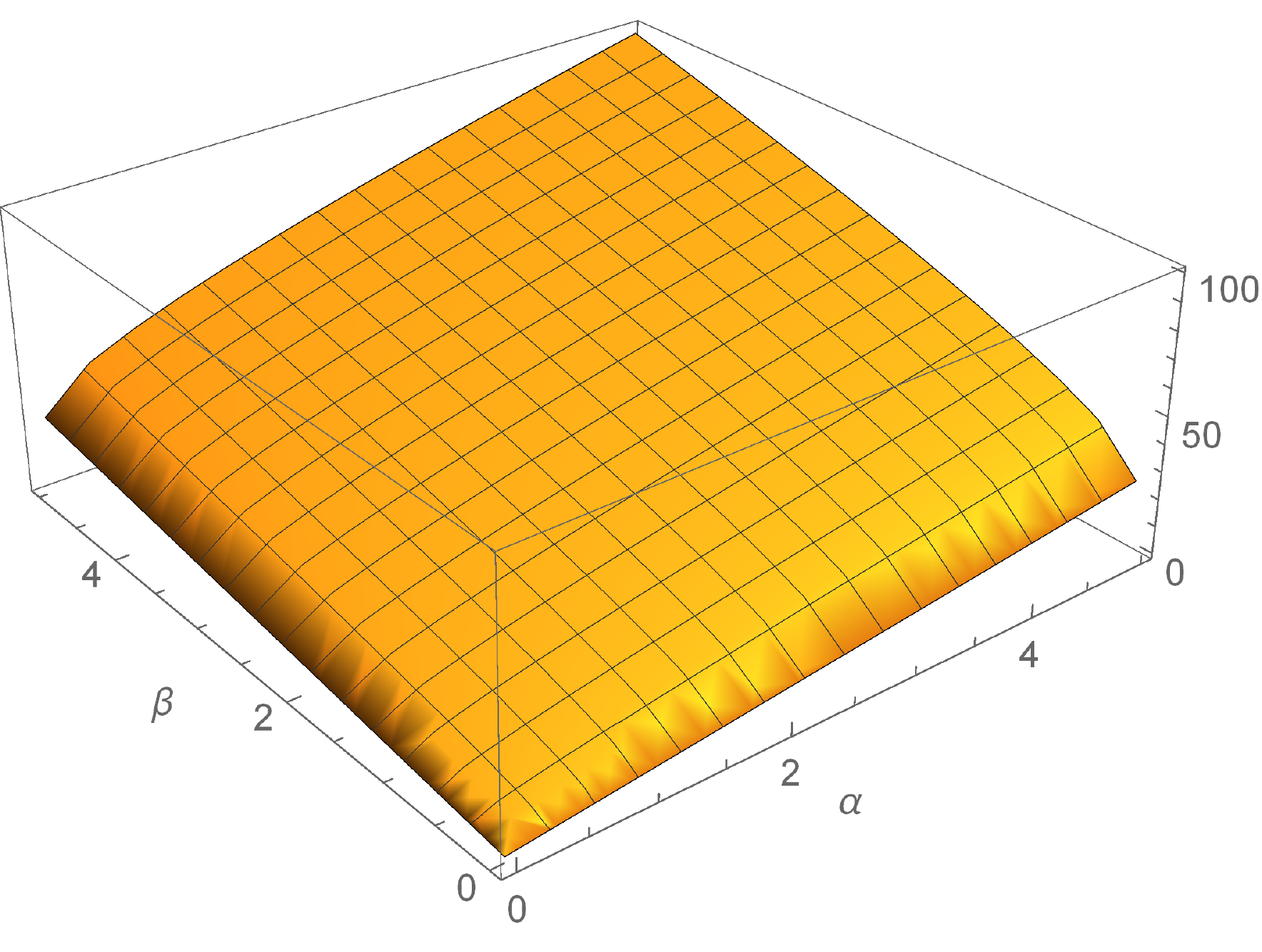}\tabularnewline
{\footnotesize{}(a) $\underline{\psi}$} & {\footnotesize{}(b) $\overline{\psi}=\overline{\varphi}$}\tabularnewline
\end{tabular}

\begin{tabular}{c}
\includegraphics[width=0.45\textwidth]{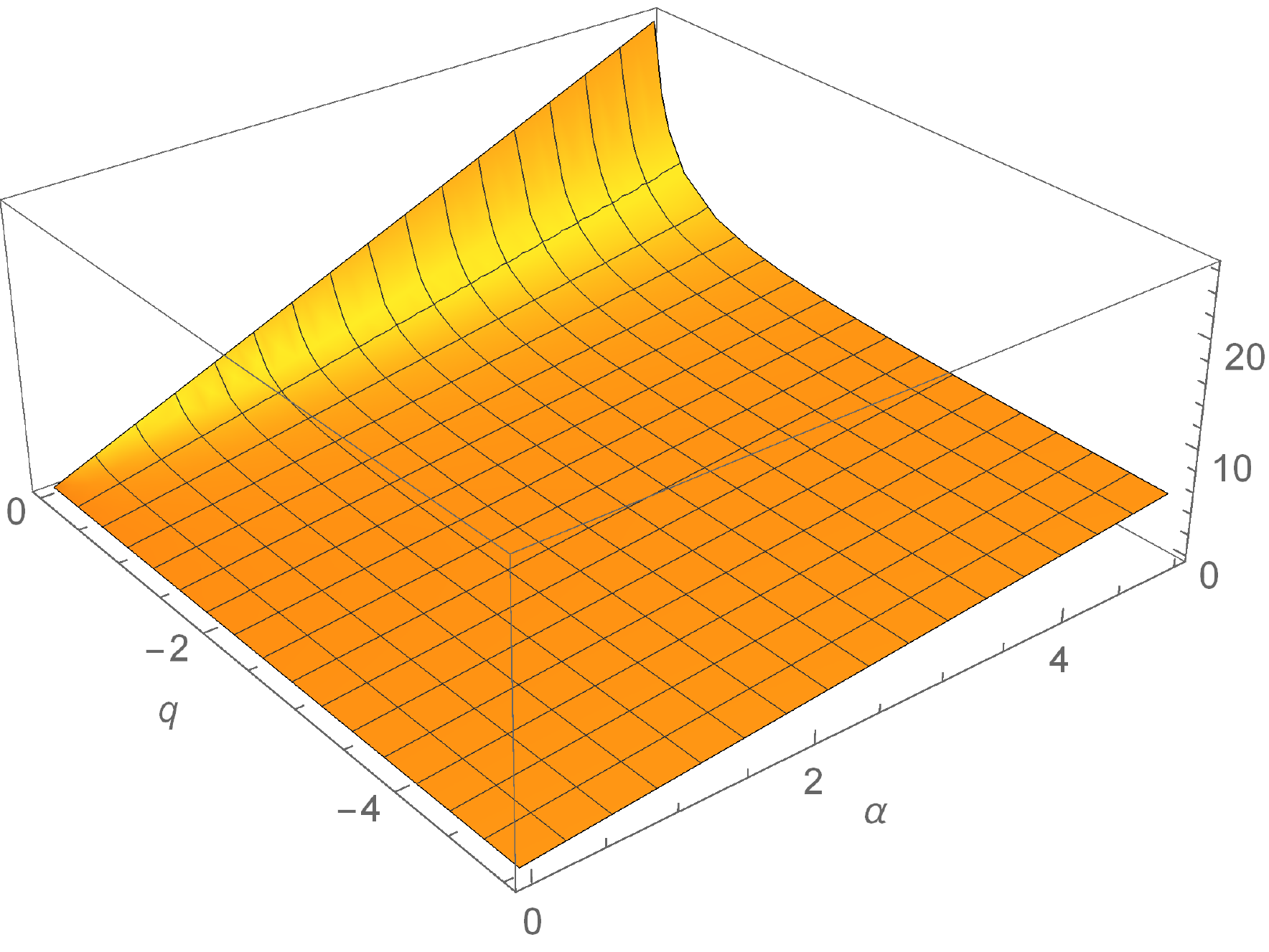}\tabularnewline
{\footnotesize{}(c) $\varphi_{q}$}\tabularnewline
\end{tabular}

\caption{\label{fig:phiGaussian}Illustration of $\underline{\psi}$, $\overline{\psi}=\overline{\varphi}$,
and $\varphi_{q}$ for the bivariate Gaussian source with the correlation
coefficient $\rho=0.9$. Lemma \ref{lem:convexity-Gaussian} implies
that $\underline{\psi}$ is convex, $\overline{\psi}=\overline{\varphi}$
is concave, and $\varphi_{q}$ is linear for $q<0$. In fact, both
the graphs of $\underline{\psi}$ and $\overline{\psi}=\overline{\varphi}$
consist of half lines emanating from the origin.}
\end{figure*}

As for the Gaussian source, the analytic expressions for envelopes
are given in the following lemma, which is a consequence of classic
hypercontractivity inequalities. See details in Appendix \ref{sec:Proof-of-Lemma}. 
\begin{lem}
\label{lem:convexity-Gaussian} For the bivariate Gaussian source
$\mathcal{N}(\boldsymbol{0},\boldsymbol{\Sigma})$ with $\boldsymbol{\Sigma}=\begin{bmatrix}1 & \rho\\
\rho & 1
\end{bmatrix}$ and $\rho\in(0,1)$, the following hold. 
\begin{enumerate}
\item It holds that for $\alpha,\beta\ge0$, 
\begin{align}
 & \conv\underline{\varphi}(\alpha,\beta)\nonumber \\
 & =\underline{\psi}(\alpha,\beta)\nonumber \\
 & =\begin{cases}
\underline{\varphi}(\alpha,\beta)=\frac{\alpha+\beta-2\rho\sqrt{\alpha\beta}}{1-\rho{}^{2}}, & \rho^{2}\alpha\le\beta\le\frac{\alpha}{\rho^{2}}\\
\alpha, & \beta<\rho^{2}\alpha\\
\beta, & \alpha<\rho^{2}\beta
\end{cases}.\label{eq:exprphi-2}
\end{align}
Moreover, they are convex on $[0,\infty)^{2}$.  An optimal distribution
attaining $\underline{\varphi}(\alpha,\beta)$ for the case $\rho^{2}\alpha\le\beta\le\frac{\alpha}{\rho^{2}}$
is $Q_{XY}=\mathcal{N}((a,b),\boldsymbol{\Sigma})$, where $a=\sqrt{2\alpha},b=\sqrt{2\beta}$. 
\item It holds that 
\[
\overline{\varphi}(\alpha,\beta)=\frac{\alpha+\beta+2\rho\sqrt{\alpha\beta}}{1-\rho{}^{2}},
\]
which is increasing in one argument given the other one. Moreover,
$\overline{\varphi}$ is strictly concave on $[0,\infty)^{2}$. 
\item It holds that for $q<0$, 
\[
\varphi_{q}(\alpha)=\frac{(1-q)\alpha}{1-q-\rho^{2}},
\]
which is increasing and linear on $[0,\infty)$. 
\end{enumerate}
\end{lem}
The functions appearing in Lemma \ref{lem:convexity-Gaussian} are
plotted in Fig. \ref{fig:phiGaussian}. 

\section{\label{sec:Proof-of-Theorem-mi}Proof of Theorem \ref{thm:mi}}

Proof of $\Upsilon(\alpha,\beta)\le\Upsilon^{*}(\alpha,\beta)$: We
denote $U,V$ both following $\mathrm{Bern}(\frac{1}{2})$ such that
$X\stackrel{\BSC(a)}{\longrightarrow}U\stackrel{\BSC(c)}{\longrightarrow}V\stackrel{\BSC(b)}{\longrightarrow}Y$,
where $a*b*c=p$. Such $(a,b,c)$ exists if $a*b\le p$. If we set
$W=(U,V)$, then 
\begin{align*}
I(X,Y;W) & =1+h(p)-h(a)-h(b)\\
I(X;W) & =1-h(a)\\
I(Y;W) & =1-h(b).
\end{align*}
This leads to the desired result for $(\alpha,\beta)\in\mathcal{D}_{2}$.

For the third clause, we set $W\sim\mathrm{Bern}(\frac{1}{2})$ such
that $W\stackrel{\BSC(a)}{\longrightarrow}X\stackrel{\BSC(p)}{\longrightarrow}Y$.
For such $W$, we have 
\begin{align*}
I(X,Y;W) & =1-h(a)\\
I(X;W) & =1-h(a)\\
I(Y;W) & =1-h\left(a*p\right).
\end{align*}
If $a*p\le b$, then $I(Y;W)\ge\beta$, i.e., this $W$ is feasible.
This leads to the desired result for $(\alpha,\beta)\in\mathcal{D}_{3}$,
and by symmetry, also leads to the one for $(\alpha,\beta)\in\mathcal{D}_{4}$. 

For $(\alpha,\beta)\in\mathcal{D}_{1}$, we set $W\sim\mathrm{Bern}(\frac{1}{2})$
such that $W\stackrel{\BSC(a)}{\longrightarrow}X$ and $W\stackrel{\BSC(b)}{\longrightarrow}Y$,
and moreover, $P_{XY|W}$ is a coupling of two channels $\BSC(a)$
and $\BSC(b)$, given by 
\begin{align*}
P_{XY|W=0} & =\begin{bmatrix}1-\frac{a+b+p}{2} & \frac{-a+b+p}{2}\\
\frac{a-b+p}{2} & \frac{a+b-p}{2}
\end{bmatrix},\\
P_{XY|W=1} & =\begin{bmatrix}\frac{a+b-p}{2} & \frac{a-b+p}{2}\\
\frac{-a+b+p}{2} & 1-\frac{a+b+p}{2}
\end{bmatrix}.
\end{align*}
For such $(W,X,Y)$, the marginal distribution on $(X,Y)$ is 
\[
P_{XY}=\frac{1}{2}P_{XY|W=0}+\frac{1}{2}P_{XY|W=1}=\begin{bmatrix}\frac{1-p}{2} & \frac{p}{2}\\
\frac{p}{2} & \frac{1-p}{2}
\end{bmatrix},
\]
which coincides with the given distribution. So, such $W$ is feasible.
This leads to the desired result for $(\alpha,\beta)\in\mathcal{D}_{1}$. 
\begin{center}
\par\end{center}

Proof of $\Upsilon(\alpha,\beta)\ge\Upsilon^{*}(\alpha,\beta)$: Observe
that 
\begin{align*}
 & \Upsilon(\alpha,\beta)-\alpha-\beta\\
 & \ge\inf_{P_{W|XY}:I(X;W)\ge\alpha,I(Y;W)\ge\beta}I(X,Y;W)\\
 & \qquad-I(X;W)-I(Y;W)\\
 & \ge\inf_{P_{W|XY}}I(X,Y;W)-I(X;W)-I(Y;W)\\
 & =\inf_{P_{W|XY}}-I(X;Y)+I(X;Y|W)\\
 & \ge-I(X;Y).
\end{align*}
Hence, 
\begin{align*}
\Upsilon(\alpha,\beta) & \ge\alpha+\beta-I(X;Y)\\
 & =1+h(p)-h(a)-h(b).
\end{align*}
This implies the desired result for $(\alpha,\beta)\in\mathcal{D}_{2}$. 

Furthermore, 
\begin{align*}
 & \Upsilon(\alpha,\beta)-\alpha\\
 & \ge\inf_{P_{W|XY}:I(X;W)\ge\alpha,I(Y;W)\ge\beta}I(X,Y;W)-I(X;W)\\
 & \ge0.
\end{align*}
Hence, $\Upsilon(\alpha,\beta)\ge\alpha=1-h(a)$, which implies the
desired result for $(\alpha,\beta)\in\mathcal{D}_{3}$, and by symmetry,
also implies the one for $(\alpha,\beta)\in\mathcal{D}_{4}$. 

We now consider $(\alpha,\beta)\in\mathcal{D}_{1}$. Observe that
for any $R_{XY}$, 
\begin{align}
\Upsilon(\alpha,\beta) & =\inf_{P_{W|XY}:I(X;W)\ge\alpha,I(Y;W)\ge\beta}I(X,Y;W)\nonumber \\
 & =\inf_{\substack{Q_{WXY}:Q_{XY}=P_{XY},\\
D(Q_{X|W}\|R_{X}|Q_{W})-D(P_{X}\|R_{X})\ge\alpha,\\
D(Q_{Y|W}\|R_{Y}|Q_{W})-D(P_{Y}\|R_{Y})\ge\beta
}
}\nonumber \\
 & \qquad D(Q_{XY|W}\|R_{XY}|Q_{W})-D(P_{XY}\|R_{XY})\\
 & \ge\inf_{\substack{Q_{WXY}:\\
D(Q_{X|W}\|R_{X}|Q_{W})-D(P_{X}\|R_{X})\ge\alpha,\\
D(Q_{Y|W}\|R_{Y}|Q_{W})-D(P_{Y}\|R_{Y})\ge\beta
}
}\nonumber \\
 & \qquad D(Q_{XY|W}\|R_{XY}|Q_{W})-D(P_{XY}\|R_{XY})\\
 & =\conv\underline{\psi}(\alpha+D(P_{X}\|R_{X}),\beta+D(P_{Y}\|R_{Y}))\nonumber \\
 & \qquad-D(P_{XY}\|R_{XY})\label{eq:-31}\\
 & =\underline{\psi}(\alpha+D(P_{X}\|R_{X}),\beta+D(P_{Y}\|R_{Y}))\\
 & \qquad-D(P_{XY}\|R_{XY}),\label{eq:}
\end{align}
where $\underline{\psi}$ is defined in \eqref{eq:lce-1} but for
$R_{XY}$, \eqref{eq:-31} follows by Lemma \ref{lem:timesharing}
(recall that $\conv\underline{\psi}$ denotes the lower convex envelope
of $\underline{\psi}$), and the last line follows by the convexity
of $\underline{\psi}$ shown in Statement 3 of Lemma \ref{lem:convexity}.

We now choose $R_{XY}=\DSBS(\hat{p})$ with $\hat{p}\in(0,1/2)$.
The value of $\hat{p}$ will be specified later. For such $R_{XY}$,
it holds that 
\begin{align*}
D(P_{X}\|R_{X}) & =D(P_{Y}\|R_{Y})=0.
\end{align*}
Moreover, by Lemma \ref{lem:convexity} again, for the case of $b\le a*\hat{p},a\le b*\hat{p}$,
it holds that 
\begin{align*}
\underline{\psi}(\alpha,\beta) & =\underline{\varphi}(\alpha,\beta)\\
 & =D\left(\begin{bmatrix}1+q-a-b & b-q\\
a-q & q
\end{bmatrix}\|\begin{bmatrix}\frac{1-\hat{p}}{2} & \frac{\hat{p}}{2}\\
\frac{\hat{p}}{2} & \frac{1-\hat{p}}{2}
\end{bmatrix}\right),
\end{align*}
 where $a=h^{-1}(1-\alpha),b=h^{-1}(1-\beta)$, and 
\begin{align*}
q & =q_{a,b}(\hat{p}):=\frac{1}{2\left(\kappa-1\right)}\times\Bigl(\left(\kappa-1\right)\left(a+b\right)+1\\
 & \qquad\qquad-\sqrt{\left(\left(\kappa-1\right)\left(a+b\right)+1\right)^{2}-4\kappa\left(\kappa-1\right)ab}\Bigr)
\end{align*}
with $\kappa=\left(\frac{1-\hat{p}}{\hat{p}}\right)^{2}$.  

 Under the condition that $a\le b$, the conditions that $b\le a*\hat{p},a\le b*\hat{p}$
are equivalent to $\hat{p}\ge\frac{b-a}{1-2a}$. Observe that $q_{a,b}(\hat{p})$
is continuous in $\hat{p}$. Moreover,  by definition, it is easily
verified that 
\begin{align*}
\lim_{\hat{p}\uparrow1/2}q_{a,b}(\hat{p}) & =ab,\\
\lim_{\hat{p}\downarrow\frac{b-a}{1-2a}}q_{a,b}(\hat{\rho}) & =\frac{a(1-a-b)}{1-2a}.
\end{align*}
On the other hand, for the case of $a*p>b,a*b>p$, it holds that
$ab<\frac{a+b-p}{2}<\frac{a(1-a-b)}{1-2a}$. So, there is a $\hat{p}^{*}\in(\frac{b-a}{1-2a},1/2)$
such that $q_{a,b}(\hat{p}^{*})=\frac{a+b-p}{2}$.  For such $\hat{p}^{*}$,
the optimal distribution $Q_{XY}$ attaining $\underline{\varphi}(\alpha,\beta)$
with $R_{XY}=\DSBS(\hat{p}^{*})$ is 
\[
Q_{XY}=\begin{bmatrix}1-\frac{a+b+p}{2} & \frac{-a+b+p}{2}\\
\frac{a-b+p}{2} & \frac{a+b-p}{2}
\end{bmatrix}.
\]

We choose $\hat{p}=\hat{p}^{*}$, i.e., $R_{XY}=\DSBS(\hat{p}^{*})$.
We then obtain that for $(\alpha,\beta)\in\mathcal{D}_{1}$, 
\begin{align*}
\Upsilon(\alpha,\beta) & \ge\underline{\varphi}(\alpha,\beta)-D(P_{XY}\|R_{XY})\\
 & =D(Q_{XY}\|R_{XY})-D(P_{XY}\|R_{XY})\\
 & =-H_{Q}(X,Y)-\mathbb{E}_{Q}\log R_{XY}(X,Y)\\
 & \qquad+H_{P}(X,Y)-\mathbb{E}_{P}\log R_{XY}(X,Y)\\
 & =H_{P}(X,Y)-H_{Q}(X,Y)\\
 & =1+h\left(p\right)-H\left(\begin{bmatrix}1-\frac{a+b+p}{2} & \frac{-a+b+p}{2}\\
\frac{a-b+p}{2} & \frac{a+b-p}{2}
\end{bmatrix}\right).
\end{align*}
The last line is exactly the expression for $(\alpha,\beta)\in\mathcal{D}_{1}$.
This proves the desired result for $(\alpha,\beta)\in\mathcal{D}_{1}$.
 We hence complete the proof.

\section{\label{sec:Proof-of-Theorem-mi-1}Proof of Theorem \ref{thm:mi2}}

\subsection{Proof of Statement 2}

We first prove Statement 2. That is, for $(\alpha,\beta)\in\mathcal{I}_{0}^{*}$,
\begin{align}
\underline{\Upsilon}(\alpha,\beta) & =\underline{\Upsilon}^{*}(\alpha,\beta),\label{eq:-2}\\
\overline{\Upsilon}(\alpha,\beta) & =\overline{\Upsilon}^{*}(\alpha,\beta).\label{eq:-3}
\end{align}
Note that here $(\alpha,\beta)\in\mathcal{I}_{0}^{*}$ instead of
$(\alpha,\beta)\in\mathcal{I}_{0}$.

\textbf{Proof of \eqref{eq:-2}:} We first consider the equality in
\eqref{eq:-2}. By definition, $\underline{\Upsilon}(\alpha,\beta)\ge\Upsilon(\alpha,\beta)$.
Moreover, for $(\alpha,\beta)\in\mathcal{D}_{1}\cup\mathcal{D}_{2}$,
$\Upsilon^{*}(\alpha,\beta)=\underline{\Upsilon}^{*}(\alpha,\beta)$.
So, by Theorem \ref{thm:mi}, $\underline{\Upsilon}(\alpha,\beta)\ge\Upsilon(\alpha,\beta)=\Upsilon^{*}(\alpha,\beta)=\underline{\Upsilon}^{*}(\alpha,\beta)$.
On the other hand, the random variable $W$ constructed in the proof
of Theorem \ref{thm:mi} in fact satisfies $I(X;W)=\alpha,I(Y;W)=\beta$,
and $I(X,Y;W)=\Upsilon^{*}(\alpha,\beta)=\underline{\Upsilon}^{*}(\alpha,\beta)$.
So, $\underline{\Upsilon}(\alpha,\beta)=\underline{\Upsilon}^{*}(\alpha,\beta)$
for $(\alpha,\beta)\in\mathcal{D}_{1}\cup\mathcal{D}_{2}$.  We next
consider $(\alpha,\beta)\in\mathcal{D}_{3}'\cup\mathcal{D}_{3}''\cup\mathcal{D}_{4}'\cup\mathcal{D}_{4}''$. 

By replacing the inequality constraints in the infimizations with
the corresponding equality constraints in the equation chain in \eqref{eq:},
it holds that for any $R_{XY}$, 
\begin{align}
\underline{\Upsilon}(\alpha,\beta) & \ge\conv\underline{\varphi}(\alpha+D(P_{X}\|R_{X}),\beta+D(P_{Y}\|R_{Y}))\nonumber \\
 & \qquad-D(P_{XY}\|R_{XY}),\label{eq:-1}
\end{align}
where $\underline{\varphi}$ is defined in \eqref{eq:-20} but for
$R_{XY}$. We now choose $R_{XY}=\DSBS(\hat{p})$ with $\hat{p}>0$.
For such $R_{XY}$, it holds that 
\begin{align*}
D(P_{X}\|R_{X}) & =D(P_{Y}\|R_{Y})=0.
\end{align*}

We now consider the case $(\alpha,\beta)\in\mathcal{D}_{3}'$, i.e.,
$h^{-1}(1-\beta/\alpha)\le p<\frac{b-a}{1-2a}$. For this case, we
choose $\hat{p}$ in the same range $h^{-1}(1-\beta/\alpha)\le\hat{p}<\frac{b-a}{1-2a}$.
For this case, we choose $\hat{p}\le h^{-1}(1-\beta/\alpha)$. By
Statement 4 in Lemma \ref{lem:convexity} (more precisely, by Remark
\ref{rem:optimaldistribution}), the optimal distribution $Q_{WXY}$
attaining $\conv\underline{\varphi}(\alpha,\beta)$ is given by $Q_{W}=\mathrm{Bern}(\theta)$,
and 
\begin{align*}
Q_{XY|W=0} & =\begin{bmatrix}\frac{1-\hat{p}}{2} & \frac{\hat{p}}{2}\\
\frac{\hat{p}}{2} & \frac{1-\hat{p}}{2}
\end{bmatrix},\\
Q_{XY|W=1} & =\begin{bmatrix}(1-a')\left(1-\hat{p}\right) & (1-a')\hat{p}\\
a'\hat{p} & a'\left(1-\hat{p}\right)
\end{bmatrix}.
\end{align*}
where $a'\in[0,1/2]$ is the unique solution to the equation $\frac{1-h(a'*\hat{p})}{1-h(a')}=\frac{\beta}{\alpha}$,
 and $\theta=\frac{\alpha}{1-h(a')}$. This distribution satisfies
that 
\begin{align*}
D(Q_{X|W}\|R_{X}) & =\alpha,\\
D(Q_{Y|W}\|R_{Y}) & =\beta,\\
D(Q_{XY|W}\|R_{XY}|Q_{W}) & =\conv\underline{\varphi}(\alpha,\beta).
\end{align*}
For such a distribution, 
\[
Q_{XY}(0,1)+Q_{XY}(1,0)=(1-\theta)\hat{p}+\theta\hat{p}=\hat{p}.
\]

We choose $\hat{p}=p$.  Substituting such a choice of $\hat{p}$
into the inequality in \eqref{eq:-1} yields that 
\begin{align*}
\underline{\Upsilon}(\alpha,\beta) & \ge\conv\underline{\varphi}(\alpha,\beta)-D(P_{XY}\|R_{XY})\\
 & =D(Q_{XY|W}\|R_{XY}|Q_{W})-D(P_{XY}\|R_{XY})\\
 & =-H_{Q}(X,Y|W)-\mathbb{E}_{Q}\log R_{XY}(X,Y)\\
 & \qquad+H_{P}(X,Y)-\mathbb{E}_{P}\log R_{XY}(X,Y)\\
 & =H_{P}(X,Y)-H_{Q}(X,Y|W)\\
 & =1+h(p)-(1-\theta)\left(1+h(p)\right)\\
 & \qquad-\theta H\left(\begin{bmatrix}(1-a')\left(1-p\right) & (1-a')p\\
a'p & a'\left(1-p\right)
\end{bmatrix}\right)\\
 & =1+h(p)-(1-\theta)\left(1+h(p)\right)\\
 & \qquad-\theta\left(h(a')+h(p)\right)\\
 & =\theta\left(1-h(a')\right)\\
 & =\alpha.
\end{align*}
This completes the proof of the case $(\alpha,\beta)\in\mathcal{D}_{3}'$.
By symmetry, the desired result still holds for $(\alpha,\beta)\in\mathcal{D}_{4}'.$

We next consider the case $(\alpha,\beta)\in\mathcal{D}_{3}''$, i.e.,
$\alpha h^{-1}(1-\beta/\alpha)\le p\le h^{-1}(1-\beta/\alpha)$. For
this case, we choose $\hat{p}$ such that $\hat{p}\le h^{-1}(1-\beta/\alpha)$.
By Statement 4 in Lemma \ref{lem:convexity} (more precisely, by
Remark \ref{rem:optimaldistribution}), the optimal distribution $Q_{WXY}$
attaining $\conv\underline{\varphi}(\alpha,\beta)$ is given by $Q_{W}=\mathrm{Bern}(\alpha)$,
and 
\begin{align*}
Q_{XY|W=0} & =\begin{bmatrix}\frac{1-\hat{p}}{2} & \frac{\hat{p}}{2}\\
\frac{\hat{p}}{2} & \frac{1-\hat{p}}{2}
\end{bmatrix},\\
Q_{XY|W=1} & =\begin{bmatrix}1-h^{-1}(1-\beta/\alpha) & h^{-1}(1-\beta/\alpha)\\
0 & 0
\end{bmatrix}.
\end{align*}
This distribution satisfies that 
\begin{align*}
D(Q_{X|W}\|R_{X}) & =\alpha,\\
D(Q_{Y|W}\|R_{Y}) & =\beta,\\
D(Q_{XY|W}\|R_{XY}|Q_{W}) & =\conv\underline{\varphi}(\alpha,\beta).
\end{align*}
For such a distribution, 
\[
Q_{XY}(0,1)+Q_{XY}(1,0)=(1-\alpha)\hat{p}+\alpha h^{-1}(1-\beta/\alpha).
\]

We choose $\hat{p}\in[0,h^{-1}(1-\beta/\alpha)]$ such that 
\[
\hat{p}\left(1-\alpha\right)+\alpha h^{-1}(1-\beta/\alpha)=p.
\]
Such $\hat{p}$ always exists for the case of $(\alpha,\beta)\in\mathcal{D}_{3}''$,
since for this case, $\alpha h^{-1}(1-\beta/\alpha)\le p\le h^{-1}(1-\beta/\alpha)$.

Substituting such a choice of $\hat{p}$ into the inequality in \eqref{eq:-1}
yields that 
\begin{align*}
 & \underline{\Upsilon}(\alpha,\beta)\\
 & \ge\conv\underline{\varphi}(\alpha,\beta)-D(P_{XY}\|R_{XY})\\
 & =D(Q_{XY|W}\|R_{XY}|Q_{W})-D(P_{XY}\|R_{XY})\\
 & =-H_{Q}(X,Y|W)-\mathbb{E}_{Q}\log R_{XY}(X,Y)\\
 & \qquad+H_{P}(X,Y)-\mathbb{E}_{P}\log R_{XY}(X,Y)\\
 & =H_{P}(X,Y)-H_{Q}(X,Y|W)\\
 & =1+h(p)-(1-\alpha)\left(1+h(\hat{p})\right)-\alpha\left(1-\beta/\alpha\right)\\
 & =h(p)+\beta-(1-\alpha)h\left(\frac{p-\alpha h^{-1}(1-\beta/\alpha)}{1-\alpha}\right).
\end{align*}
This completes the proof of the case $(\alpha,\beta)\in\mathcal{D}_{3}''$.
By symmetry, the desired result still holds for $(\alpha,\beta)\in\mathcal{D}_{4}''.$

\textbf{Proof of \eqref{eq:-3}:} We next prove the equality in \eqref{eq:-3}.
On one hand, 

\begin{align*}
\overline{\Upsilon}(\alpha,\beta) & =\sup_{\substack{P_{W|XY}:I(X;W)=\alpha,\\
I(Y;W)=\beta
}
}I(X,Y;W)\\
 & =\sup_{\substack{P_{W|XY}:I(X;W)=\alpha,\\
I(Y;W)=\beta
}
}H(X,Y)-H(X,Y|W)\\
 & \le\sup_{\substack{P_{W|XY}:I(X;W)=\alpha,\\
I(Y;W)=\beta
}
}H(X,Y)-H(Y|W)\\
 & =\sup_{\substack{P_{W|XY}:I(X;W)=\alpha,\\
I(Y;W)=\beta
}
}H(X|Y)+I(Y;W)\\
 & =h(p)+\beta.
\end{align*}
By symmetry, $\overline{\Upsilon}(\alpha,\beta)\le h(p)+\alpha.$
That is, $\overline{\Upsilon}(\alpha,\beta)\le h(p)+\alpha\land\beta=\overline{\Upsilon}^{*}(\alpha,\beta).$ 

We set $W'\stackrel{\BSC(a)}{\longrightarrow}X\stackrel{\BSC(p)}{\longrightarrow}Y$,
or equivalently, $X=W'\oplus Z',\;Y=X\oplus Z$ where $W'\sim\mathrm{Bern}(\frac{1}{2})$,
$Z'\sim\mathrm{Bern}(a),$ and $Z\sim\mathrm{Bern}(p)$ are mutually
independent. Here $\oplus$ denotes the XOR operation (i.e., the module-2
sum). Set $W=(W',Z)$. For such $W$, we have 
\begin{align*}
I(X,Y;W) & =I(X;W)+I(Y;W|X)\\
 & =I(X;W)+H(Y|X)\\
 & =1-h(a)+h(p),\\
I(X;W) & =I(X;W',Z)=I(X;W')\\
 & =1-h(a),\\
I(Y;W) & =I(Y;W',Z)=1-H(Y|W',Z)\\
 & =1-H(Z'|W',Z)=1-H(Z')\\
 & =1-h(a).
\end{align*}
So, 
\begin{equation}
\overline{\Upsilon}(\alpha,\alpha)\ge h(p)+\alpha=\overline{\Upsilon}^{*}(\alpha,\alpha).\label{eq:-4}
\end{equation}
Moreover, from the expression derived for the lower envelope, we observe
that for $\beta=\alpha-\alpha h(p/\alpha)$, 
\begin{align}
\overline{\Upsilon}(\alpha,\beta)\ge\underline{\Upsilon}(\alpha,\beta) & =\underline{\Upsilon}^{*}(\alpha,\beta)=h(p)+\beta,\label{eq:-5}
\end{align}
and for $\alpha=\beta-\beta h(p/\beta)$, 
\begin{align}
\overline{\Upsilon}(\alpha,\beta)\ge\underline{\Upsilon}(\alpha,\beta) & =\underline{\Upsilon}^{*}(\alpha,\beta)=h(p)+\alpha.\label{eq:-6}
\end{align}
Combining \eqref{eq:-4}-\eqref{eq:-6} yields $\overline{\Upsilon}(\alpha,\beta)\ge h(p)+\alpha\land\beta=\overline{\Upsilon}^{*}(\alpha,\beta).$
Hence, $\overline{\Upsilon}(\alpha,\beta)=\overline{\Upsilon}^{*}(\alpha,\beta)$
for $(\alpha,\beta)\in\mathcal{I}_{0}^{*}$. 

\subsection{Proof of Statement 1}

Observe that the upper envelope $\overline{\Upsilon}$ and the lower
envelope $\underline{\Upsilon}$ coincide on the curves $\beta=\alpha-\alpha h(p/\alpha)$
and $\alpha=\beta-\beta h(p/\beta)$. By the monotonicity of $\overline{\Upsilon}$
and $\underline{\Upsilon}$, the projection region $\mathcal{I}_{0}$
must be exactly $\mathcal{I}_{0}^{*}$, since, otherwise, $\overline{\Upsilon}<\underline{\Upsilon}$
holds on the region $\mathcal{I}_{0}\backslash\mathcal{I}_{0}^{*}$
which contradicts with the obvious fact that $\overline{\Upsilon}\ge\underline{\Upsilon}$.

\section{\label{sec:Proof-of-Theorem-Gaussian}Proof of Theorem \ref{thm:miGaussian} }

We first prove Statement 2. We now consider the equality in \eqref{eq:mi-7}.
 We first prove the ``$\le$'' part. We denote $W\sim\mathcal{N}(0,1)$
and $(\hat{X},\hat{Y})\sim\mathcal{N}((a,b),\hat{\boldsymbol{\Sigma}})$
with 
\[
\hat{\boldsymbol{\Sigma}}=\begin{bmatrix}1 & \hat{\rho}\\
\hat{\rho} & 1
\end{bmatrix}.
\]
Assume $W$ and $(\hat{X},\hat{Y})$ are independent. Denote 
\begin{align}
X & =\sqrt{1-N_{1}}W+\sqrt{N_{1}}\hat{X}\label{eq:-12}\\
Y & =\sqrt{1-N_{2}}W+\sqrt{N_{2}}\hat{Y}\label{eq:-13}
\end{align}
with $N_{1},N_{2}\in[0,1]$. Then the correlation coefficient between
$X$ and $Y$ is 
\begin{align*}
 & \mathbb{E}\left[(\sqrt{1-N_{1}}W+\sqrt{N_{1}}\hat{X})(\sqrt{1-N_{2}}W+\sqrt{N_{2}}\hat{Y})\right]\\
 & =\sqrt{\left(1-N_{1}\right)\left(1-N_{2}\right)}+\hat{\rho}\sqrt{N_{1}N_{2}}.
\end{align*}

We choose $\hat{\rho}\in[0,1]$ such that the resultant correlation
coefficient is exactly $\rho$, i.e., 
\[
\hat{\rho}=\frac{\rho-\sqrt{\left(1-N_{1}\right)\left(1-N_{2}\right)}}{\sqrt{N_{1}N_{2}}}.
\]
For this case, the joint distribution of $X,Y$ is exactly $P_{XY}$.
The induced mutual informations are respectively 
\begin{align*}
I(X,Y;W) & =\frac{1}{2}\ln\left(\frac{1-\rho^{2}}{N_{1}N_{2}\left(1-\hat{\rho}^{2}\right)}\right)\\
I(X;W) & =\frac{1}{2}\ln\frac{1}{N_{1}}\\
I(Y;W) & =\frac{1}{2}\ln\frac{1}{N_{2}}.
\end{align*}
Let $I(X;W)=\alpha,I(Y;W)=\beta$. Then, $N_{1}=e^{-2\alpha},N_{2}=e^{-2\beta}$,
which then implies 
\begin{align*}
\hat{\rho} & =\rho_{\alpha,\beta},\\
I(X,Y;W) & =\alpha+\beta+\frac{1}{2}\ln\left(\frac{1-\rho^{2}}{1-\hat{\rho}^{2}}\right).
\end{align*}
Recall the definition of $\rho_{\alpha,\beta}$ in \eqref{eq:rhohat}.
Therefore, 
\begin{equation}
\underline{\Upsilon}(\alpha,\beta)\le\alpha+\beta+\frac{1}{2}\ln\left(\frac{1-\rho^{2}}{1-\hat{\rho}_{\alpha,\beta}^{2}}\right)\label{eq:-11}
\end{equation}
as long as $0\le\rho_{\alpha,\beta}<1$. Hence, this inequality holds
for $(\alpha,\beta)\in\mathcal{D}_{\mathrm{G},1}$. 

For $(\alpha,\beta)\in\mathcal{D}_{\mathrm{G},2}$, we choose $\hat{\rho}=0$,
which leads to $\underline{\Upsilon}(\alpha,\beta)\le\alpha+\beta-\frac{1}{2}\ln\frac{1}{1-\rho^{2}}$. 

For $(\alpha,\beta)\in\mathcal{D}_{\mathrm{G},3}$, although the bound
in \eqref{eq:-11} still holds, we can derive a better bound by choosing
a better $W$. Note that the curve $\left\{ (\alpha,\beta):\rho^{2}=\frac{1-e^{-2\beta}}{1-e^{-2\alpha}}\right\} \subseteq\mathcal{D}_{\mathrm{G},1}$
since $\beta\le\alpha$ and $\rho_{\alpha,\beta}=e^{\beta-\alpha}\rho\in(0,1)$
for $(\alpha,\beta)$ on the curve. So, as proven above, $\underline{\Upsilon}(\alpha,\beta)=\alpha$
on this curve. Rewrite the curve equation as $\beta=f(\alpha):=-\frac{1}{2}\ln\left(1-\rho^{2}+\rho^{2}e^{-2\alpha}\right)$,
and note that the derivative of $f$ is 
\[
f'(\alpha)=-\frac{\rho^{2}e^{-2\alpha}}{1-\rho^{2}+\rho^{2}e^{-2\alpha}}
\]
which decreases from $\rho^{2}$ to $0$ as $\alpha$ increases from
$0$ to $+\infty$.   Hence, the closed convex hull of the curve
$\left\{ \left(\alpha,f(\alpha)\right):\alpha\ge0\right\} $ (i.e.,
the graph of $f$) is the set $\left\{ \left(\alpha,\beta\right):0\le\beta\le f(\alpha),\alpha\ge0\right\} $.
By convex combination of points on the curve, we obtain that $\underline{\Upsilon}(\alpha,\beta)=\alpha$
on the set $\left\{ \left(\alpha,\beta\right):0\le\beta\le f(\alpha),\alpha\ge0\right\} $,
i.e., on $\mathcal{D}_{\mathrm{G},3}$. By symmetry, $\underline{\Upsilon}(\alpha,\beta)=\beta$
on $\mathcal{D}_{\mathrm{G},4}$.

We next prove the other direction, i.e., the ``$\ge$'' part. Observe
that for any $R_{XY}$, 
\begin{align}
\underline{\Upsilon}(\alpha,\beta) & =\inf_{P_{W|XY}:I(X;W)=\alpha,I(Y;W)=\beta}I(X,Y;W)\nonumber \\
 & =\inf_{\substack{Q_{WXY}:Q_{XY}=P_{XY},\\
D(Q_{X|W}\|R_{X}|Q_{W})-D(P_{X}\|R_{X})=\alpha,\\
D(Q_{Y|W}\|R_{Y}|Q_{W})-D(P_{Y}\|R_{Y})=\beta
}
}\nonumber \\
 & \qquad D(Q_{XY|W}\|R_{XY}|Q_{W})-D(P_{XY}\|R_{XY})\\
 & \ge\inf_{\substack{Q_{WXY}:\\
D(Q_{X|W}\|R_{X}|Q_{W})-D(P_{X}\|R_{X})=\alpha,\\
D(Q_{Y|W}\|R_{Y}|Q_{W})-D(P_{Y}\|R_{Y})=\beta
}
}\nonumber \\
 & \qquad D(Q_{XY|W}\|R_{XY}|Q_{W})-D(P_{XY}\|R_{XY})\\
 & =\conv\underline{\varphi}(\alpha',\beta')-D(P_{XY}\|R_{XY})\nonumber \\
 & =\underline{\psi}(\alpha',\beta')-D(P_{XY}\|R_{XY}),\label{eq:-17}
\end{align}
where $\alpha':=\alpha+D(P_{X}\|R_{X}),\beta':=\beta+D(P_{Y}\|R_{Y})$,
and $\underline{\varphi}$ and $\underline{\psi}$ are defined in
\eqref{eq:-20} and \eqref{eq:lce-1} but for $R_{XY}$. The last
line above follows by Statement 1 of Lemma \ref{lem:convexity-Gaussian}. 

We now choose $R_{XY}=\mathcal{N}(\boldsymbol{0},\boldsymbol{\Sigma}_{R})$
with $\boldsymbol{\Sigma}_{R}=\begin{bmatrix}N_{1} & \hat{\rho}\sqrt{N_{1}N_{2}}\\
\hat{\rho}\sqrt{N_{1}N_{2}} & N_{2}
\end{bmatrix}$, $N_{1}=e^{-2\alpha},N_{2}=e^{-2\beta}$, and $\hat{\rho}\ge0$.
  For such $R_{XY}$, it holds that
\begin{align}
D(P_{X}\|R_{X}) & =\frac{1}{2}\left(\ln N_{1}+\frac{1}{N_{1}}-1\right)\nonumber \\
D(P_{Y}\|R_{Y}) & =\frac{1}{2}\left(\ln N_{2}+\frac{1}{N_{2}}-1\right)\nonumber \\
D(P_{XY}\|R_{XY}) & =\frac{1}{2}\left(\ln\frac{|\boldsymbol{\Sigma}_{R}|}{|\boldsymbol{\Sigma}|}+\tr\left(\boldsymbol{\Sigma}_{R}^{-1}\boldsymbol{\Sigma}\right)-2\right)\nonumber \\
 & =\frac{1}{2}\Bigl(\ln\frac{N_{1}N_{2}\left(1-\hat{\rho}^{2}\right)}{1-\rho^{2}}\nonumber \\
 & \quad+\frac{N_{1}+N_{2}-2\rho\hat{\rho}\sqrt{N_{1}N_{2}}}{N_{1}N_{2}\left(1-\hat{\rho}^{2}\right)}-2\Bigr).\label{eq:-15}
\end{align}
So, 
\begin{align*}
\alpha' & =\frac{1}{2}\left(e^{2\alpha}-1\right)=\frac{1}{2}\left(\frac{1}{N_{1}}-1\right),\\
\beta' & =\frac{1}{2}\left(e^{2\beta}-1\right)=\frac{1}{2}\left(\frac{1}{N_{2}}-1\right).
\end{align*}

We now consider the case of $(\alpha,\beta)\in\mathcal{D}_{\mathrm{G},1}$.
For this case, we choose $\hat{\rho}\ge0$ such that $\hat{\rho}^{2}\alpha'\le\beta'\le\frac{\alpha'}{\hat{\rho}^{2}}$
(the specific value of $\hat{\rho}$ will be given below), and by
Statement 1 of Lemma \ref{lem:convexity-Gaussian}, an optimal distribution
attaining $\underline{\psi}(\alpha',\beta')$ (or $\underline{\varphi}(\alpha',\beta')$)
is $Q_{XY}=\mathcal{N}((a,b),\boldsymbol{\Sigma}_{R})$ with $a=\sqrt{2\alpha'N_{1}},b=\sqrt{2\beta'N_{2}}$.
 The value of $\underline{\psi}(\alpha',\beta')$ is 
\begin{align}
 & \underline{\psi}(\alpha',\beta')\nonumber \\
 & =D(Q_{XY}\|R_{XY})\\
 & =\frac{1}{2}\left(\frac{a^{2}N_{2}+b^{2}N_{1}-2ab\hat{\rho}\sqrt{N_{1}N_{2}}}{N_{1}N_{2}\left(1-\hat{\rho}{}^{2}\right)}\right)\\
 & =\frac{1}{2N_{1}N_{2}\left(1-\hat{\rho}{}^{2}\right)}\Bigl((1-N_{1})N_{2}+(1-N_{2})N_{1}\nonumber \\
 & \qquad-2\hat{\rho}\sqrt{N_{1}N_{2}\left(1-N_{1}\right)\left(1-N_{2}\right)}\Bigr).\label{eq:-14}
\end{align}

We choose  
\begin{align}
\hat{\rho} & =\rho_{\alpha,\beta},\label{eq:-16}
\end{align}
with $\rho_{\alpha,\beta}$ defined in \eqref{eq:rhohat}, which satisfies
$0\le\rho_{\alpha,\beta}<1$ for $(\alpha,\beta)\in\mathcal{D}_{\mathrm{G},1}$;
see the argument below Theorem \ref{thm:miGaussian}. Such $\hat{\rho}$
also satisfies $\hat{\rho}^{2}\alpha'\le\beta'\le\frac{\alpha'}{\hat{\rho}^{2}}$,
i.e., $\hat{\rho}\le\min\left\{ \sqrt{\frac{\alpha'}{\beta'}},\sqrt{\frac{\beta'}{\alpha'}}\right\} $,
as desired, since this condition is equivalent to that 
\[
\rho\le\min\left\{ \frac{\cos\theta_{\beta}}{\cos\theta_{\alpha}},\frac{\cos\theta_{\alpha}}{\cos\theta_{\beta}}\right\} .
\]
 Substituting \eqref{eq:-15}, \eqref{eq:-14}, and \eqref{eq:-16}
into \eqref{eq:-17} yields that 
\begin{align*}
\underline{\Upsilon}(\alpha,\beta) & \ge\alpha+\beta+\frac{1}{2}\ln\left(\frac{1-\rho^{2}}{1-\rho_{\alpha,\beta}^{2}}\right).
\end{align*}

We now consider the case $(\alpha,\beta)\in\mathcal{D}_{\mathrm{G},2}$
in which $\rho_{\alpha,\beta}\le0$. For this case, we choose $\hat{\rho}=0$
which yields 
\begin{align*}
\underline{\Upsilon}(\alpha,\beta) & \ge\alpha+\beta+\frac{1}{2}\ln\left(1-\rho^{2}\right).
\end{align*}

We now consider the case of $(\alpha,\beta)\in\mathcal{D}_{\mathrm{G},3}$.
For this case, we choose $\hat{\rho}=\rho$, i.e., $R_{XY}=P_{XY}$.
For this case, by Statement 1 of Lemma \ref{lem:convexity-Gaussian},
it holds that $\underline{\Upsilon}(\alpha,\beta)=\underline{\psi}(\alpha,\beta)\ge\alpha$,
since for this case, $\rho>\frac{\cos\theta_{\beta}}{\cos\theta_{\alpha}}=\sqrt{\frac{1-e^{-2\beta}}{1-e^{-2\alpha}}}\ge\sqrt{\frac{\beta}{\alpha}}$
by the facts that $t\mapsto\frac{1-e^{-2t}}{t}$ is decreasing and
 $\beta<\alpha$. By symmetry, $\underline{\Upsilon}(\alpha,\beta)\ge\beta$
for $(\alpha,\beta)\in\mathcal{D}_{\mathrm{G},4}$. 

Combining all cases above, \eqref{eq:mi-7} holds. 

The finiteness of $\underline{\Upsilon}(\alpha,\beta)$ on $[0,\infty)^{2}$
implies the projection region $\mathcal{I}_{0}=[0,\infty)^{2},$i.e.,
Statement 1.

We next prove \eqref{eq:mi-8}. Denote 
\begin{align}
X & =\sqrt{1-N}\hat{W}+\sqrt{N}\hat{X}\label{eq:-12-1}\\
Y & =\rho X+\sqrt{1-\rho^{2}}\hat{Y}\label{eq:-13-1}
\end{align}
with $N\in(0,1)$ and independent standard Gaussian random variables
$\hat{W},\hat{X},\hat{Y}$. Denote $W=(\hat{W},\hat{Y})$. For this
case, 
\begin{align*}
I(X;W) & =I(X;\hat{W})\\
 & =\frac{1}{2}\ln\frac{1}{N}<+\infty,\\
I(Y;W) & =h_{\mathrm{d}}(Y)-h_{\mathrm{d}}(Y|\hat{W},\hat{Y})\\
 & =\frac{1}{2}\ln\frac{1}{N\rho^{2}}<+\infty,\\
I(X,Y;W) & =I(X;W)+I(Y;W|X)\\
 & =\frac{1}{2}\ln\frac{1}{N}+h_{\mathrm{d}}(Y|X)-h_{\mathrm{d}}(Y|X,\hat{W},\hat{Y})\\
 & =+\infty,
\end{align*}
where $h_{\mathrm{d}}$ denotes the differential entropy. Hence, $\overline{\Upsilon}(\alpha,\beta)=+\infty$
for some finite point $(\alpha,\beta)$. By the concavity of $\overline{\Upsilon}$,
it holds that $\overline{\Upsilon}(\alpha,\beta)=+\infty$ for all
$\alpha,\beta\ge0$, i.e., \eqref{eq:mi-8}.

\appendices{}

\section{\label{sec:Proof-of-Lemma-1}Proofs of Statements 2 and 4 in Lemma
\ref{lem:convexity}}

\subsection{Proof of Statement 2 in Lemma \ref{lem:convexity}}

The monotonicity in fact follows by the convexity of the relative
entropy. Specifically, by the strict convexity of the relative entropy,
it holds that $(a,b)\mapsto\D((1-a,a),(1-b,b)\|P_{XY})$ is strictly
convex, since for $Q_{XY}^{(i)}$ attaining $\D((1-a_{i},a_{i}),(1-b_{i},b_{i})\|P_{XY})$,
$i=0,1$, 
\begin{align*}
 & (1-\lambda)D(Q_{XY}^{(0)}\|P_{XY})+\lambda D(Q_{XY}^{(1)}\|P_{XY})\\
 & >D((1-\lambda)Q_{XY}^{(0)}+\lambda Q_{XY}^{(1)}\|P_{XY})\\
 & \ge\D((1-a_{\lambda},a_{\lambda}),(1-b_{\lambda},b_{\lambda})\|P_{XY}),
\end{align*}
where $a_{\lambda}=(1-\lambda)a_{0}+\lambda a_{1},\:b_{\lambda}=(1-\lambda)b_{0}+\lambda b_{1}$.
So, given $\alpha\in[0,1]$ (or equivalently, given $a=h^{-1}(1-\alpha)\in[0,1/2]$),
the function $b\mapsto\D((1-a,a),(1-b,b)\|P_{XY})$ is strictly convex,
and its minimum is $\alpha$ which is attained at $b=a*p$. On the
other hand, observe that 
\begin{align}
 & \inf_{\hat{\beta}\ge\beta}\underline{\varphi}(\alpha,\hat{\beta})\nonumber \\
 & =\inf_{\hat{b}\in[0,1/2]:1-h(\hat{b})\ge\beta}\D((1-a,a),(1-\hat{b},\hat{b})\|P_{XY})\\
 & =\inf_{\hat{b}\in[0,b]}\D((1-a,a),(1-\hat{b},\hat{b})\|P_{XY}).\label{eq:-30}
\end{align}
By the strict convexity of the objective function at the last line,
the minimum is uniquely attained by $\hat{b}=b$ when $b\le a*p$.
Hence, $\underline{\varphi}(\alpha,\hat{\beta})>\underline{\varphi}(\alpha,\beta)$
for all $\hat{\beta},\beta$ such that $\hat{\beta}>\beta$ and $b\le a*p$.
That is, $\beta\mapsto\underline{\varphi}(\alpha,\beta)$ is strictly
decreasing for $\beta$ such that $a*p\le b$. 

The strict monotonicity of $\beta\mapsto\underline{\varphi}(\alpha,\beta)$
on the interval $a*p\ge b$ can be proven similarly (by replacing
the constraints $\hat{\beta}\ge\beta$, $1-h(\hat{b})\ge\beta$, and
$\hat{b}\in[0,b]$ in \eqref{eq:-30} respectively with $\hat{\beta}\le\beta$,
$1-h(\hat{b})\le\beta$, and $\hat{b}\in[b,1/2]$). 

\subsection{Proof of Statement 4 in Lemma \ref{lem:convexity}}

We first make the following claim. 

\textbf{Claim 1:} $\conv\underline{\varphi}(\alpha,\beta)=\underline{\psi}(\alpha,\beta)$
for $(\alpha,\beta)$ such that $\beta\ge\left(1-h(p)\right)\alpha$
and $\alpha\ge\left(1-h(p)\right)\beta$. In other words, the formula
in \eqref{eq:exprphi-3} holds for $(\alpha,\beta)\in\hat{\mathcal{D}}_{1}\cup\hat{\mathcal{D}}_{2}\cup\hat{\mathcal{D}}_{3}$.

We now prove this claim. On one hand, by Statement 3 and the definition
of $\underline{\psi}$, 
\begin{equation}
\conv\underline{\varphi}(\alpha,\beta)\ge\conv\underline{\psi}(\alpha,\beta)=\underline{\psi}(\alpha,\beta).\label{eq:-29}
\end{equation}
On the other hand, $\conv\underline{\varphi}(\alpha,\beta)\le\underline{\varphi}(\alpha,\beta)=\underline{\psi}(\alpha,\beta)$
for $(\alpha,\beta)$ such that $a*p\ge b,\;b*p\ge a$ (i.e., $(\alpha,\beta)\in\hat{\mathcal{D}}_{1}$).
So, \eqref{eq:exprphi-3} holds for $(\alpha,\beta)\in\hat{\mathcal{D}}_{1}$. 

Denote $\beta^{*}(\alpha)$ as the value $\beta$ such that $a*p=b$
where $a=h^{-1}(1-\alpha),b=h^{-1}(1-\beta)$. By definition of $\underline{\varphi}$,
it is easily verified that $\underline{\varphi}(\alpha,\beta^{*}(\alpha))=\alpha$,
and hence, all points $(\alpha,\beta^{*}(\alpha),\underline{\varphi}(\alpha,\beta^{*}(\alpha)))$
with $\alpha\in[0,1]$ are coplanar. That is, they are on the plane
$\left\{ (\alpha,\beta,\alpha):\alpha,\beta\in[0,1]\right\} $. So,
$\conv\underline{\varphi}(\alpha,\beta)\le\alpha$ for $(\alpha,\beta)$
such that $a*p\le b,\,\beta\ge\left(1-h(p)\right)\alpha$ (i.e., $(\alpha,\beta)\in\hat{\mathcal{D}}_{2}$;
see this region in the subfigure (a) in Fig. \ref{fig:phi}). Since
\eqref{eq:-29} still holds and $\underline{\psi}(\alpha,\beta)=\alpha$
for this case (by Statement 3), it holds that $\conv\underline{\varphi}(\alpha,\beta)=\alpha$
for $(\alpha,\beta)\in\hat{\mathcal{D}}_{2}$. Similarly, $\conv\underline{\varphi}(\alpha,\beta)=\beta$
for  $(\alpha,\beta)\in\hat{\mathcal{D}}_{3}$. This completes the
proof of the claim above.

We next consider the case $(\alpha,\beta)\in\hat{\mathcal{D}}_{4}$.
 By Statement 2, given $\alpha$, $\beta\mapsto\underline{\varphi}(\alpha,\beta)$
is strictly decreasing for $a*p\le b$ (and hence also for $\beta<\left(1-h(p)\right)\alpha$).
Based on these observations, if we denote $(1/u,1/v)$ as a subgradient
of $\conv\underline{\varphi}$ at $(\alpha,\beta)$ with $\beta<\left(1-h(p)\right)\alpha$,
then $v<0$. 

If $(\alpha_{i},\beta_{i},\underline{\varphi}(\alpha_{i},\beta_{i})),i\in[3]$
are on the supporting plane of $\conv\underline{\varphi}$ at $(\alpha,\beta)$,
then $(\alpha_{i},\beta_{i}),i\in[3]$ must attain the following minimum:,
\[
\Gamma:=\min_{s,t\ge0}\underline{\varphi}(s,t)-\frac{s}{u}-\frac{t}{v}.
\]
We now make the second claim. 

\textbf{Claim 2:} Any optimal $(s^{*},t^{*})$ attaining the minimum
above must be either $(0,0)$ or $(1,\beta')$ for some $\beta'\in[0,1]$. 

We next prove this claim. By the definition of $\varphi_{q}$ in \eqref{eq:phi_q-2},
we can rewrite 

\[
\Gamma=\min_{s\ge0}\varphi_{v}(s)-\frac{s}{u}.
\]
By Statement 5 in Lemma \ref{lem:convexity}, for $v<0$, $\varphi_{v}$
is strictly concave on $[0,1]$. So, the infimum above is only attained
at $s=0$ or $1$. Moreover, for $s=0$, it holds that 
\begin{align*}
\varphi_{v}(0) & =\min_{0\le t\le1}\underline{\varphi}(0,t)-\frac{t}{v}=0,
\end{align*}
since $\underline{\varphi}(0,t)-\frac{t}{v}\ge0$ (note $v<0$) and
this lower bound is uniquely attained at $t=0$. So, the unique minimizer
above is $t=0$. For $s=1$, it holds that 
\begin{align*}
\varphi_{v}(1) & =\min_{0\le t\le1}\underline{\varphi}(1,t)-\frac{t}{v}\\
 & =\min_{0\le b\le1}\D((1,0),(1-b,b)\|P_{XY})\\
 & \qquad-\frac{D((1-b,b)\|P_{Y})}{v}\\
 & =\min_{0\le b\le1}D\left(\begin{bmatrix}1-b & b\\
0 & 0
\end{bmatrix}\|P_{XY}\right)\\
 & \qquad-\frac{D((1-b,b)\|P_{Y})}{v}\\
 & =\min_{0\le b\le1}\left(\frac{1}{v}-1\right)h(b)-\left(1-b\right)\log\left(1-\frac{p}{2}\right)\\
 & \qquad-b\log\frac{p}{2}-\frac{1}{v}.
\end{align*}
Since $v<0$, it holds that the objective function in the last line
is strictly convex, the minimum is attained by a unique $b$. This
completes the proof of Claim 2. 

By Claim 2, $(\alpha,\beta,\conv\underline{\varphi}(\alpha,\beta))$
is the convex combination of $(0,0,0)$ and $(1,\beta',\underline{\varphi}(1,\beta'))$
for some $\beta'\in[0,1]$ (see the subfigure (a) in Fig. \ref{fig:phi}
for better understanding this statement). That is, 
\[
(\alpha,\beta,\conv\underline{\varphi}(\alpha,\beta))=(1-\theta)(0,0,0)+\theta(1,\beta',\underline{\varphi}(1,\beta')),
\]
which implies 
\begin{align*}
\beta' & =\beta/\alpha\\
\theta & =\alpha.
\end{align*}
These parameters induce the following optimal distribution $Q_{WXY}$
which attains $\conv\underline{\varphi}(\alpha,\beta)$. Here $W$
denotes the time-sharing (or convex-combination) variable. The optimal
distribution $Q_{WXY}$ is given by $Q_{W}=\mathrm{Bern}(\alpha)$,
and 
\begin{align*}
Q_{XY|W=0} & =\begin{bmatrix}\frac{1-p}{2} & \frac{p}{2}\\
\frac{p}{2} & \frac{1-p}{2}
\end{bmatrix},\\
Q_{XY|W=1} & =\begin{bmatrix}1-h^{-1}(1-\beta/\alpha) & h^{-1}(1-\beta/\alpha)\\
0 & 0
\end{bmatrix}.
\end{align*}
Hence, for this case, 
\begin{align*}
\conv\underline{\varphi}(\alpha,\beta) & =D(Q_{XY|W}\|R_{XY}|Q_{W})\\
 & =\alpha+\alpha D\bigl((1-h^{-1}(1-\beta/\alpha),\\
 & \qquad h^{-1}(1-\beta/\alpha))\|(1-p,p)\bigr).
\end{align*}
This proves \eqref{eq:exprphi-3} for $(\alpha,\beta)\in\hat{\mathcal{D}}_{4}$.
The case $(\alpha,\beta)\in\hat{\mathcal{D}}_{5}$ follows by symmetry. 

\section{\label{sec:Proof-of-Lemma}Proof of Lemma \ref{lem:convexity-Gaussian}}

 The {\em forward} and {\em reverse hypercontractivity regions}
 for a joint distribution $P_{XY}$ are respectively 
\begin{align}
\mathcal{R}_{\mathrm{FH}}(P_{XY}) & :=\big\{(p,q)\in[1,\infty)^{2}:\nonumber \\
 & \qquad\langle f,g\rangle\le\Vert f\Vert_{p}\Vert g\Vert_{q},~\forall\,f,g\ge0\big\}\label{eq:FIFHR}
\end{align}
and 
\begin{align}
\mathcal{R}_{\mathrm{RH}}(P_{XY}) & :=\big\{(p,q)\in(-\infty,1]^{2}:\nonumber \\
 & \qquad\langle f,g\rangle\ge\Vert f\Vert_{p}\Vert g\Vert_{q},~\forall\,f,g\ge0\big\},\label{eq:FIRHR}
\end{align}
where $f:\mathcal{X}\to[0,\infty)$ and $g:\mathcal{Y}\to[0,\infty)$
denote nonnegative measurable functions, $\langle f,g\rangle:=\mathbb{E}_{P}\left[f(X)g(Y)\right]$
denote the inner product of $f$ and $g$, and $\Vert f\Vert_{p}:=\mathbb{E}_{P}\left[f^{p}(X)\right]^{1/p}$
and $\Vert g\Vert_{q}:=\mathbb{E}_{P}\left[g^{q}(Y)\right]^{1/q}$
are respectively the (pseudo) $p$-norm of $f$ and the (pseudo) $q$-norm
of $g$. In other words, the forward and reverse hypercontractivity
regions are respectively the sets of parameters $(p,q)$ such that
the forward and reverse hypercontractivity inequalities hold. 

We can write $\mathcal{R}_{\mathrm{RH}}(P_{XY})$ as the disjoint
union of four sets 
\begin{align}
\mathcal{R}_{\mathrm{RH}}^{++}(P_{XY}) & :=(0,1]^{2}\cap\mathcal{R}_{\mathrm{RH}}(\pi_{XY}),\label{eqn:subreg1}\\
\mathcal{R}_{\mathrm{RH}}^{+-}(P_{XY}) & :=\big((0,1]\times(-\infty,0)\big)\cap\mathcal{R}_{\mathrm{RH}}(\pi_{XY}),\label{eqn:subreg2}\\
\mathcal{R}_{\mathrm{RH}}^{-+}(P_{XY}) & :=\big((-\infty,0)\times(0,1]\big)\cap\mathcal{R}_{\mathrm{RH}}(\pi_{XY}),\label{eqn:subreg3}\\
\mathcal{R}_{\mathrm{RH}}^{--}(P_{XY}) & :=(-\infty,0]^{2}.\label{eqn:subreg4}
\end{align}
The forward and reverse hypercontractivity regions admit the information-theoretic
characterizations  \cite{ahlswede1976spreading,carlen2009subadditivity,nair2014equivalent,kamath2015reverse,beigi2016equivalent,liu2018information,yu2021strong}:
\begin{align}
\mathcal{R}_{\mathrm{FH}}(P_{XY}) & =\Big\{(p,q)\in[1,\infty)^{2}:\nonumber \\
 & \qquad\underline{\psi}(\alpha,\beta)\ge\frac{\alpha}{p}+\frac{\beta}{q},\,\forall\,\alpha,\beta\ge0\Big\},\label{eq:FIFHR-2}\\
\mathcal{R}_{\mathrm{RH}}^{++}(P_{XY}) & =\Big\{(p,q)\in(0,1]^{2}:\nonumber \\
 & \qquad\overline{\varphi}(\alpha,\beta)\le\frac{\alpha}{p}+\frac{\beta}{q},\,\forall\,\alpha,\beta\ge0\Big\},\\
\mathcal{R}_{\mathrm{RH}}^{+-}(P_{XY}) & =\Big\{(p,q)\in(0,1]\times(-\infty,0):\nonumber \\
 & \qquad\varphi_{q}(\alpha)\leq\frac{\alpha}{p},\,\forall\,\alpha\ge0\Big\}.\label{eq:-7}
\end{align}
By symmetry, $\mathcal{R}_{\mathrm{RH}}^{-+}(P_{XY})$ can be characterized
in an analogous manner to $\mathcal{R}_{\mathrm{RH}}^{+-}(P_{XY})$
in \eqref{eq:-7}. 

Furthermore, for the bivariate Gaussian source $P_{XY}$ with correlation
coefficient $\rho\in(0,1)$, it is well known (e.g., \cite{ODonnell14analysisof})
that the forward and reverse hypercontractivity regions are respectively
explicitly given by 
\begin{align*}
\mathcal{R}_{\mathrm{FH}}(P_{XY}) & =\big\{(p,q)\in[1,\infty)^{2}:(p-1)(q-1)\ge\rho^{2}\big\},\\
\mathcal{R}_{\mathrm{RH}}(P_{XY}) & =\big\{(p,q)\in(-\infty,1]^{2}:(p-1)(q-1)\ge\rho^{2}\big\}.
\end{align*}
Therefore, by the information-theoretic characterizations above, for
such a source, 
\begin{align}
\underline{\psi}(\alpha,\beta) & \ge\sup_{(p,q)\in[1,\infty)^{2}:(p-1)(q-1)\ge\rho^{2}}\frac{\alpha}{p}+\frac{\beta}{q}\label{eq:-8}\\
 & =\begin{cases}
\frac{\alpha+\beta-2\rho\sqrt{\alpha\beta}}{1-\rho{}^{2}} & \rho^{2}\alpha\le\beta\le\frac{\alpha}{\rho^{2}}\\
\alpha & \beta<\rho^{2}\alpha\\
\beta & \beta>\frac{\alpha}{\rho^{2}}
\end{cases},\\
\overline{\varphi}(\alpha,\beta) & \le\inf_{(p,q)\in(0,1]^{2}:(p-1)(q-1)\ge\rho^{2}}\frac{\alpha}{p}+\frac{\beta}{q}\label{eq:-10}\\
 & =\frac{\alpha+\beta+2\rho\sqrt{\alpha\beta}}{1-\rho{}^{2}},\\
\varphi_{q}(\alpha) & \leq\inf_{p\in(0,1]:(p-1)(q-1)\ge\rho^{2}}\frac{\alpha}{p}\label{eq:-9}\\
 & =\frac{(1-q)\alpha}{1-q-\rho^{2}}\textrm{ for }q<0.\label{eq:-32}
\end{align}
The optimal choice of $(p,q)$ attaining the supremum in \eqref{eq:-8}
is $p=\frac{(1-\rho^{2})\sqrt{\alpha}}{\sqrt{\alpha}-\rho\sqrt{\beta}},q=\frac{(1-\rho^{2})\sqrt{\beta}}{\sqrt{\beta}-\rho\sqrt{\alpha}}$
for the case of $\rho^{2}\alpha<\beta<\frac{\alpha}{\rho^{2}}$. The
optimal choice of $(p,q)$ attaining the infimum in \eqref{eq:-10}
is $p=\frac{(1-\rho^{2})\sqrt{\alpha}}{\sqrt{\alpha}+\rho\sqrt{\beta}},q=\frac{(1-\rho^{2})\sqrt{\beta}}{\sqrt{\beta}+\rho\sqrt{\alpha}}$
for all $\alpha,\beta\ge0$. The optimal choice of $p$ attaining
the infimum in \eqref{eq:-9} is $p=1+\frac{\rho^{2}}{1-q}$ for all
$\alpha\ge0$. 

We now prove that the inequalities in \eqref{eq:-8}-\eqref{eq:-32}
are in fact equalities. For the Gaussian source $P_{XY}=\mathcal{N}(\boldsymbol{0},\boldsymbol{\Sigma})$
with $\boldsymbol{\Sigma}=\begin{bmatrix}1 & \rho\\
\rho & 1
\end{bmatrix}$, if we choose $Q_{X}=\mathcal{N}(a,1)$ and $Q_{Y}=\mathcal{N}(b,1)$,
then 
\begin{align}
D(Q_{X}\|P_{X}) & =\frac{a^{2}}{2}\label{eq:-23}\\
D(Q_{Y}\|P_{Y}) & =\frac{b^{2}}{2}\nonumber \\
\D(Q_{X},Q_{Y}\|P_{XY}) & =\frac{1}{2}\left(\frac{a^{2}+b^{2}-2\rho ab}{1-\rho{}^{2}}\right).\label{eq:-24}
\end{align}
The last equality above follows since 
\begin{align*}
 & \D(Q_{X},Q_{Y}\|P_{XY})\\
 & =\inf_{Q_{XY}\in\calC(Q_{X},Q_{Y})}D(Q_{XY}\|P_{XY})\\
 & \ge\inf_{\rho'\in[0,1]}D(Q_{XY}^{(\boldsymbol{\Sigma}')}\|P_{XY})\\
 & =\inf_{\rho'\in[0,1]}D(Q_{XY}^{(\boldsymbol{\Sigma}')}\|\mathcal{N}((a,b),\boldsymbol{\Sigma}'))\\
 & \qquad+D(\mathcal{N}((a,b),\boldsymbol{\Sigma}')\|P_{XY})\\
 & \ge\inf_{\rho'\in[0,1]}D(\mathcal{N}((a,b),\boldsymbol{\Sigma}')\|P_{XY})\\
 & =D(\mathcal{N}((a,b),\boldsymbol{\Sigma})\|P_{XY})\\
 & =\frac{1}{2}\left(\frac{a^{2}+b^{2}-2\rho ab}{1-\rho{}^{2}}\right),
\end{align*}
and this lower bound is attained at $Q_{XY}=\mathcal{N}((a,b),\boldsymbol{\Sigma})$,
where $\boldsymbol{\Sigma}':=\begin{bmatrix}1 & \rho'\\
\rho' & 1
\end{bmatrix}$, and $Q_{XY}^{(\boldsymbol{\Sigma}')}$ denotes a joint distribution
with covariance matrix $\boldsymbol{\Sigma}'$.  For $\rho^{2}\alpha\le\beta\le\frac{\alpha}{\rho^{2}}$,
we choose $a=\sqrt{2\alpha},b=\sqrt{2\beta}$ and then obtain $\underline{\psi}(\alpha,\beta)\le\underline{\varphi}(\alpha,\beta)\le\frac{\alpha+\beta-2\rho\sqrt{\alpha\beta}}{1-\rho{}^{2}}$,
which, combined with \eqref{eq:-8}, implies the equality in \eqref{eq:-8}
for the case of $\rho^{2}\alpha\le\beta\le\frac{\alpha}{\rho^{2}}$
(and also $\underline{\psi}(\alpha,\beta)=\underline{\varphi}(\alpha,\beta)$
for this case). By the monotonicity, the equality in \eqref{eq:-8}
also holds for the case of $\beta<\rho^{2}\alpha$ or $\beta>\frac{\alpha}{\rho^{2}}$.
For the inequality in \eqref{eq:-10}, we choose $a=\sqrt{2\alpha},b=-\sqrt{2\beta}$
which verifies the inequality in \eqref{eq:-10}. 

Similarly, for $Q_{X}=\mathcal{N}(a,1)$, it holds that for $q<0$,
\begin{align*}
 & \inf_{Q_{Y}}\D(Q_{X},Q_{Y}\|P_{XY})-\frac{D(Q_{Y}\|P_{Y})}{q}\\
 & =\inf_{Q_{Y|X}}D(Q_{XY}\|P_{XY})-\frac{D(Q_{Y}\|P_{Y})}{q}\\
 & \ge\inf_{b\in\mathbb{R},\rho''\in[0,1],N_{2}\ge0}D(Q_{XY}^{(\boldsymbol{\Sigma}'')}\|P_{XY})-\frac{D(Q_{Y}^{(\boldsymbol{\Sigma}'')}\|P_{Y})}{q}\\
 & =\inf_{b\in\mathbb{R},\rho''\in[0,1],N_{2}\ge0}D(Q_{XY}^{(\boldsymbol{\Sigma}'')}\|\mathcal{N}((a,b),\boldsymbol{\Sigma}''))\\
 & \qquad+D(\mathcal{N}((a,b),\boldsymbol{\Sigma}'')\|P_{XY})\\
 & \qquad-\frac{D(Q_{Y}^{(\boldsymbol{\Sigma}'')}\|\mathcal{N}(b,N_{2}))+D(\mathcal{N}(b,N_{2})\|P_{Y})}{q}\\
 & \ge\inf_{b\in\mathbb{R},\rho''\in[0,1],N_{2}\ge0}D(\mathcal{N}((a,b),\boldsymbol{\Sigma}'')\|P_{XY})\\
 & \qquad-\frac{D(\mathcal{N}(b,N_{2})\|P_{Y})}{q}\\
 & =\inf_{b\in\mathbb{R}}D(\mathcal{N}((a,b),\boldsymbol{\Sigma})\|P_{XY})-\frac{D(\mathcal{N}(b,1)\|P_{Y})}{q}\\
 & =\inf_{b\in\mathbb{R}}\frac{1}{2}\left(\frac{a^{2}+b^{2}-2\rho ab}{1-\rho{}^{2}}-\frac{b^{2}}{q}\right)\\
 & =\inf_{b\in\mathbb{R}}\frac{1}{2}\left(\frac{a^{2}+b^{2}-2\rho ab}{1-\rho{}^{2}}\right),\\
 & =\frac{(1-q)a^{2}}{2(1-q-\rho^{2})},
\end{align*}
and this lower bound is attained at $Q_{XY}=\mathcal{N}((a,b),\boldsymbol{\Sigma})$
with $b=\frac{-\rho qa}{1-q-\rho^{2}}$, where $\boldsymbol{\Sigma}'':=\begin{bmatrix}1 & \rho''\sqrt{N_{2}}\\
\rho''\sqrt{N_{2}} & N_{2}
\end{bmatrix}$, and $Q_{XY}^{(\boldsymbol{\Sigma}'')}$ denotes a joint distribution
with covariance matrix $\boldsymbol{\Sigma}''$. We choose $a=\sqrt{2\alpha}$
here, which verifies the equality in \eqref{eq:-32}.

We now prove $\conv\underline{\varphi}(\alpha,\beta)=\underline{\psi}(\alpha,\beta)$.
On one hand, $\conv\underline{\varphi}(\alpha,\beta)\ge\conv\underline{\psi}(\alpha,\beta)=\underline{\psi}(\alpha,\beta)$.
On the other hand, by choosing \eqref{eq:-23}-\eqref{eq:-24}, $\conv\underline{\varphi}(\alpha,\beta)\le\underline{\varphi}(\alpha,\beta)\le\frac{\alpha+\beta-2\rho\sqrt{\alpha\beta}}{1-\rho{}^{2}}$
for $\rho^{2}\alpha\le\beta\le\frac{\alpha}{\rho^{2}}$. So, $\conv\underline{\varphi}(\alpha,\beta)=\underline{\psi}(\alpha,\beta)$
for $\rho^{2}\alpha\le\beta\le\frac{\alpha}{\rho^{2}}$. 

 For the case $\beta<\rho^{2}\alpha$, we choose $Q_{X}=\mathcal{N}(a,N)$,
and choose $Q_{Y}$ as the output distribution of channel $P_{Y|X}$
when the input distribution is $Q_{X}$. So, $Q_{Y}=\mathcal{N}(\rho a,1-\rho^{2}+\rho^{2}N)$.
For this case, 
\begin{align*}
D(Q_{X}\|P_{X}) & =\frac{1}{2}\left(\ln N+\frac{1+a^{2}}{N}-1\right),\\
D(Q_{Y}\|P_{Y}) & =\frac{1}{2}\Bigl(\ln\left(1-\rho^{2}+\rho^{2}N\right)\\
 & \qquad+\frac{1+\rho^{2}a^{2}}{1-\rho^{2}+\rho^{2}N}-1\Bigr),\\
\D(Q_{X},Q_{Y}\|P_{XY}) & =D(Q_{X}\|P_{X}).
\end{align*}
For $N\in(0,1]$ and 
\[
\beta>g(\rho):=\frac{1}{2}\left(\frac{1}{1-\rho^{2}}+\ln\left(1-\rho^{2}\right)-1\right),
\]
we choose 
\[
a=\frac{\sqrt{\left(1-\rho^{2}+\rho^{2}N\right)\left(1+2\beta-\ln\left(1-\rho^{2}+\rho^{2}N\right)\right)-1}}{\rho}
\]
which is positive and induces $D(Q_{Y}\|P_{Y})=\beta$. As $N$ decreases
from $1$ to $0$, $D(Q_{X}\|P_{X})$ increases from $\frac{\beta}{\rho^{2}}$
to $+\infty$. So, it holds that $\underline{\varphi}(\alpha,\beta)\le\alpha$
for $g(\rho)<\beta<\rho^{2}\alpha$. Note that $\underline{\varphi}(0,0)=0$.
By convex combination of $(0,0)$ and points in the region $g(\rho)<\beta<\rho^{2}\alpha$,
we obtain that $\conv\underline{\varphi}(\alpha,\beta)\le\alpha$
for $\beta<\rho^{2}\alpha$.  By symmetry, $\conv\underline{\varphi}(\alpha,\beta)\le\beta$
for $\beta>\frac{\alpha}{\rho^{2}}$. So, $\conv\underline{\varphi}(\alpha,\beta)=\underline{\psi}(\alpha,\beta)$.

\bibliographystyle{unsrt}
\bibliography{ref}

\begin{IEEEbiographynophoto}{Lei Yu} (Member, IEEE)  received the B.E. and Ph.D. degrees in electronic  
engineering from the University of Science and Technology of China (USTC)  
in 2010 and 2015, respectively. From 2015 to 2020, he worked as a 
Post-Doctoral Researcher at the USTC, National University of Singapore, and  
University of California at Berkeley. He is currently an Associate  
Professor at the School of Statistics and Data Science, LPMC, KLMDASR,  
and LEBPS, Nankai University, China. His research interests lie in the  
intersection of probability theory, information theory, and combinatorics.
\end{IEEEbiographynophoto}

\end{document}